\def\sp{\mbox {S$^+$}}
\def\spp{\mbox {S$^{++}$}}
\def\op{\mbox {O$^+$}}
\def\opp{\mbox {O$^{++}$}}
\def\teff{\mbox {$T_{\!\!\em eff}$}}
\def\tstar{\mbox {$T_*$}}
\def\qh{\mbox {$Q_{H^0}$}}
\def\x{\mbox {$R_{23}$}}
\def\oii{\mbox {[OII]\thinspace $\lambda$3727}}
\def\siii{\mbox {[SIII]\thinspace $\lambda\lambda$9069,\thinspace 9532}}

\def\msun{\mbox {${\rm M_\odot}$}}
\def\zsun{\mbox {${\rm Z_{\odot}}$}}

\def\angs{\mbox{~\AA}}
\def\line{\mbox {$\lambda$}}
\def\lines{\mbox {$\lambda\lambda$}}
\def\hii{H~II}
\def\ew{EW(H$\beta$)}
\def\halpha{H$\alpha$} 
\def\hbeta{H$\beta$}
\def\cf{{\it cf.\/\ }} 
\def\eg{{\it e.g.\/}}
 
\def\etal{{\em et\thinspace al.}\ }   

\documentstyle[11pt,aaspp4,flushrt,amssym]{article}

\lefthead{Bresolin, Kennicutt & Garnett} 
\righthead{Extragalactic HII regions}

\begin{document}

\title{\sc The ionizing stars of extragalactic \hii\/ regions
\footnotemark[1]}

\author{Fabio Bresolin\footnotemark[2], Robert C. Kennicutt, Jr.}
\affil{Steward Observatory, University of Arizona, Tucson, AZ
85721 \\ fabio@as.arizona.edu, robk@as.arizona.edu}
\and
\author{Donald R. Garnett}
\affil{Department of Astronomy, University of Minnesota, 116
Church St. SE,\\ Minneapolis, MN 55455\\ garnett@oldstyle.spa.umn.edu}

\begin{center}
{\em Accepted for publication in the Astrophysical Journal}
\end{center}

\footnotetext[1]{Observations reported here were obtained in part at
the Multiple Mirror Telescope Observatory, a joint facility of the
University of Arizona and the Smithsonian Institution.}

\footnotetext[2]{Present address: European Southern Observatory, 
Karl-Schwarzschild-Str. 2, D-85748 Garching b. M\"{u}nchen, Germany;
fbresoli@eso.org}

\begin{abstract} 
Medium-resolution spectra from 3650 \AA\/ to 10,000 \AA\/ are
presented for 96 giant \hii\/ regions distributed in 20 spiral
galaxies. In order to interpret the data, we have calculated two
separate grids of photoionization models, adopting single-star
atmospheres (Kurucz) and star clusters synthesized with different
Initial Mass Functions (IMFs) as ionizing sources.  Additional models
were computed with more recent non-LTE stellar atmospheres, in order
to check the effects of different stellar ionizing fluxes.  We use the
radiation softness parameter $\eta\prime$ =
([OII]/[OIII])/([SII]/[SIII]) of V\'{\i}lchez and Pagel to test for a
metallicity dependence of the effective temperatures of the ionizing
stars. Our results are consistent with a significant decrease in mean
stellar temperatures of the ionizing stars with increasing
metallicity.  The magnitude of the effect, combined with the behavior
of the He~I~$\lambda$5876/\hbeta\/ ratio, suggest a smaller upper mass
limit for star formation at abundances higher than solar, even when
considering the effects of metallicity on stellar evolution and
atmospheric line blanketing. However, the exact magnitudes of the
stellar temperature and IMF variations are dependent on the choice of
stellar atmosphere and evolution models used, as well as on
uncertainties in the nebular abundance scale at high metallicities.

Our results also constrain the systematic behavior of the ionization
parameter and the N/O ratio in extragalactic \hii\/ regions. The
observed spectral sequences are inconsistent with current stellar
evolution models which predict a luminous, hot W-R stellar population
in evolved \hii\/ regions older than 2--3 Myr. This suggests either
that the hardness of the emitted Lyman continuum spectrum has been
overestimated in the models, or that some mechanism disrupts the
\hii\/ regions before the W-R phases become important.
\end{abstract}

\section{INTRODUCTION}

Giant extragalactic \hii\/ regions are ionized by the massive stars of
their young embedded clusters, and as such provide powerful tools for
measuring the chemical composition of the interstellar medium and for
understanding the massive stellar populations in external galaxies
(see reviews by Shields 1990, Kennicutt 1991, Stasi\'{n}ska 1996).
Evidence for abundance gradients across the disks of spiral galaxies
was presented by Searle (1971), following pioneering work by Aller
(1942) on the radial trends of emission-line ratios. The excitation of
\hii\/ regions, measured by the ratios [OIII]\lines4959,5007/\hbeta\/ and
[OIII]\line5007/[OII]\line3727, was found to increase with
galactocentric distance, and was interpreted as an effect of the shift
from infrared fine-structure transitions to optical [OIII] emission as
the principal cooling mechanism at the lower metallicities typical of
larger radii. Later work confirmed the existence of composition
gradients by direct observation of the [OIII]\line4363 line at low
abundances (\eg\/ Smith 1975).  To explain the observed gradients in
excitation and in the equivalent width of the \hbeta\/ emission [\ew]
an additional gradient of the color temperature of the ionizing
radiation field,
\tstar, was proposed (Shields 1974, Shields \& Tinsley 1976, Shields
\& Searle 1978), with the ionizing radiation becoming harder with
increasing radius.  The work of Shields
\& Tinsley (1976) suggested further that this change in mean stellar
temperature was produced by a metallicity-dependent upper mass limit
for star formation, $M_u \propto Z^{-1/2}$. The same authors also
pointed out that a change in the temperature of the hottest stars is
to be expected, even in the absence of a metallicity effect on $M_u$,
because of opacity effects that would lower the effective temperatures
of metal-rich stars.

Subsequent efforts were concentrated on quantifying this general
picture, with the calculation of extensive grids of single-star
photoionization models.  Most authors confirmed the increase of
\tstar\/ with decreasing abundance (Stasi\'{n}ska 1980, Campbell, 
Terlevich \& Melnick 1986, V\'{\i}lchez \& Pagel 1988), together with
a possible increase in the ionization parameter (Campbell 1988). On
the other hand Evans \& Dopita (1985) argued that the available data
on the trends in the emission-line ratios were consistent with a
single ionizing temperature ($\simeq 41,500$ K) and a varying nebular
geometry and ionization parameter. As the controversy is not
completely settled, the presence of a \tstar\/ gradient still needs to
be clearly established and quantified, over the full range of
abundances observed in spirals.

The fundamental question remains whether these apparent changes in
\tstar\/ are due to systematic variations in the Initial Mass Function
(IMF) with metallicity and/or galaxy type. Abundance effects on either
the upper mass limit (Shields \& Tinsley 1976) or the slope of the IMF
(Terlevich 1985) have been proposed.  However, several other
mechanisms have been suggested, including increased stellar line
blanketing (Balick \& Sneden 1976) or internal dust absorption
(Sarazin 1976) with metallicity. Modern stellar evolution models
(\eg, Maeder 1990) predict a softening of the stellar ionizing
spectra with increasing abundance. The consequence for \hii\/ region
models is a natural decrease of \tstar, making the varying IMF
hypothesis less compelling, as shown by McGaugh (1991). Moreover, the
introduction of cluster photoionization models, in which the single
star is replaced by an evolving stellar cluster as the ionizing
source, has produced results which are consistent with a universal IMF
(Garc\'{\i}a-Vargas, Bressan \& D\'{\i}az 1995, Stasi\'{n}ska \&
Leitherer 1996).

In this paper we re-examine the evidence for \tstar\/ and IMF
gradients, using new spectra of 96 \hii\/ regions in 20 galaxies over
the 3650--10,000\angs\/ range. This work extends previous large
surveys of \hii\/ region spectra in spiral disks, including those of
McCall, Rybski \& Shields (1985), Vila-Costas \& Edmunds (1992), Oey
\& Kennicutt (1993) and Zaritsky, Kennicutt \& Huchra
(1994). The main addition of the present work is the inclusion of data
for the near-IR [SIII]\lines9069,9532 lines, which allows measurement
of the radiation softness parameter $\eta\prime$, directly related to
the ionic fraction ratio $\eta$~=~(O$^+$/O$^{++}$)/(S$^+$/S$^{++}$),
introduced by V\'{\i}lchez and Pagel (1988) as a \teff\/ sensitive
parameter. This paper extends to a large \hii\/ region sample the
$\eta\prime$ method, which so far has been applied to studies of a few
\hii\/ regions in individual galaxies. When combined with the
empirical abundance indicator \x = ([OII] + [OIII])/\hbeta\/ (Pagel
\etal 1979) and an extensive set of nebular models we can
simultaneously constrain the metallicity, the effective temperature
and the ionization parameter (D\'{\i}az \etal 1987, V\'{\i}lchez \etal
1988, D\'{\i}az \etal 1991).  A companion paper presents detailed
modeling of 40 \hii\/ regions in one galaxy (M101), with emphasis on
the \tstar\/ gradient in a homogeneous sample of \hii\/ regions
(Garnett \& Kennicutt 1998). In the present work we examine a larger
galaxy sample, spanning a wide range of abundance, with two main
goals: to study the relationship bewteen \tstar\/ and the gaseous
abundance, and to derive constraints on IMF variations with
metallicity.  For this purpose, photoionization models including as
ionizing source either an evolving star cluster or a single stellar
atmosphere are computed.  The new \hii\/ region sample is presented in
\S~2, and the nebular model grid is described in \S~3. In the
following sections we discuss our main results: systematics of \hii\/
region spectra and physical conditions (\S~4), evidence for a \tstar\/
gradient (\S~5), and consequences for the upper IMF (\S~6). We
summarize the main results in \S~7. As a quick reference a list of
abbreviations and symbols used is provided in
Table~\ref{abbreviations.table}.
 
\section{OBSERVATIONS AND DATA REDUCTION}

\subsection{Sample}
The galaxy sample (Table~\ref{galaxies.table}) spans the Hubble
sequence from type Sa to type Sm, and should therefore provide good
indication of trends of star forming region properties with
morphological type. The \hii\/ regions cover a wide range in
excitation, and their chemical abundance varies from
Z~$\simeq$~0.2~\zsun\/ to above solar (the lower limit is set by
abundance determinations found in the literature for a few of the most
metal-poor objects).  Nearly all of the observations discussed here
are new to this work, and were obtained during several runs in
1988-1992 on the Multiple Mirror Telescope (MMT) and the Steward
Observatory 2.3m Telescope; the few exceptions are indicated in
Table~\ref{galaxies.table}. The main observational material (redwards
of \hbeta) consists of spectra obtained with the MMT Red Channel
Spectrograph with an echellette grating. Most of the blue spectra
(which include the \oii\/ line) were obtained at the MMT as part of
the survey by Oey \& Kennicutt (1993). Blue data for a few individual
objects were drawn from the literature (McCall \etal 1985, Zaritsky
\etal 1994, and Dufour \etal 1980).

\subsection{MMT Echellette Spectra}
The Red Channel Spectrograph at the MMT was used in cross-dispersed
mode to obtain spectral coverage from 4700\angs\/ to 10,000\angs.
This allowed the inclusion of all the main spectral features from
\hbeta\/ to \siii. In this observing mode a 2\arcsec\/ $\times$
20\arcsec\/ slit was cross-dispersed into five orders, providing
extended wavelength coverage at a resolution of 7-15 \AA. The detector
used was a Texas Instruments thinned 800$\times$800 device which, when
binned 2$\times$2, provided a spatial scale of 0\farcs625
pixel$^{-1}$.  Exposure times ranged from 5 to 40 min.

Data reduction included a few procedures particular to the echellette
observations (see also Skillman \& Kennicutt 1993 and Kennicutt \&
Garnett 1996).  Standard star frames taken at low airmass were used to
trace the aperture for each of the five orders and to remove the
curvature introduced by the cross-disperser. Wavelength and flux
calibrations were accomplished with HeNeAr lamp frames and
observations of standard stars from Oke \& Gunn (1983) and Massey
\etal (1988), respectively. The calibrations were checked for internal
consistency using lines in the overlap regions between adjacent
orders. The residuals from the standard star fits indicate a typical
accuracy of $\pm$0.05 mag rms or better in all orders except the
reddest one ($\lambda$ $>$ 8000\angs). Here the water absorption
features in the near-IR introduce a larger uncertainty in the flux
calibration. The extinction was then modeled as a function of
wavelength by tracing the absorption features observed in the standard
star spectra. This procedure satisfactorily removes most of the
unwanted telluric features around the \siii\/ lines, as indicated by
the mean observed $\lambda$9532/$\lambda$9069 ratio = 2.47 $\pm$ 0.05
(the theoretical value is 2.44). Nevertheless we attach a conservative
uncertainty of $\pm$20\% to the [SIII] fluxes (see Kennicutt \&
Garnett 1996).  One-dimensional spectra were extracted with variable
apertures, typically eight arcseconds, but dependent on the seeing and
the size of the objects, centered on the peak of the \halpha\/
emission line. Care was taken to match as closely as possible the
regions sampled in the blue spectra.

\subsection{Blue Spectra}
Coverage of the 3650--5100\angs\/ spectral range was obtained with
two different instruments and setups. The majority of our targets in
early-type galaxies were observed with the Red Channel CCD
spectrograph on the MMT and a slit size of 2\arcsec\/ $\times$
180\arcsec\/.  These observations have been described in detail by Oey
\& Kennicutt (1993) in their project on abundances in early-type
galaxies. However, new one-dimensional extractions were made, to match
those of the red echellette spectra.

Additional objects (marked $90''$ in Table~\ref{galaxies.table}) were
observed with the B\&C spectrograph at the Steward Observatory 2.3m
telescope. Observations were made with a 600 groove mm$^{-1}$ grating
blazed at 3568\angs, providing 9\angs\/ resolution with a 4\farcs5
$\times$ 180\arcsec\/ slit. Standard reduction procedures for these
long-slit data were followed.  The \hii\/ regions in three galaxies
(M31, M33 and NGC~2403) were observed in the red using a 300 groove
mm$^{-1}$ grating blazed at 6690\angs, and providing 20\angs\/
resolution with the 4\farcs5 slit. These data provided a check on the
corresponding MMT echellette spectra.  Blue data for NGC~628,
NGC~2403, NGC~3521 and NGC~5236 were taken from the literature, with
references given in Table~\ref{galaxies.table}.

\subsection{Analysis of Spectra}
\begin{figure}[ht]
\plotfiddle{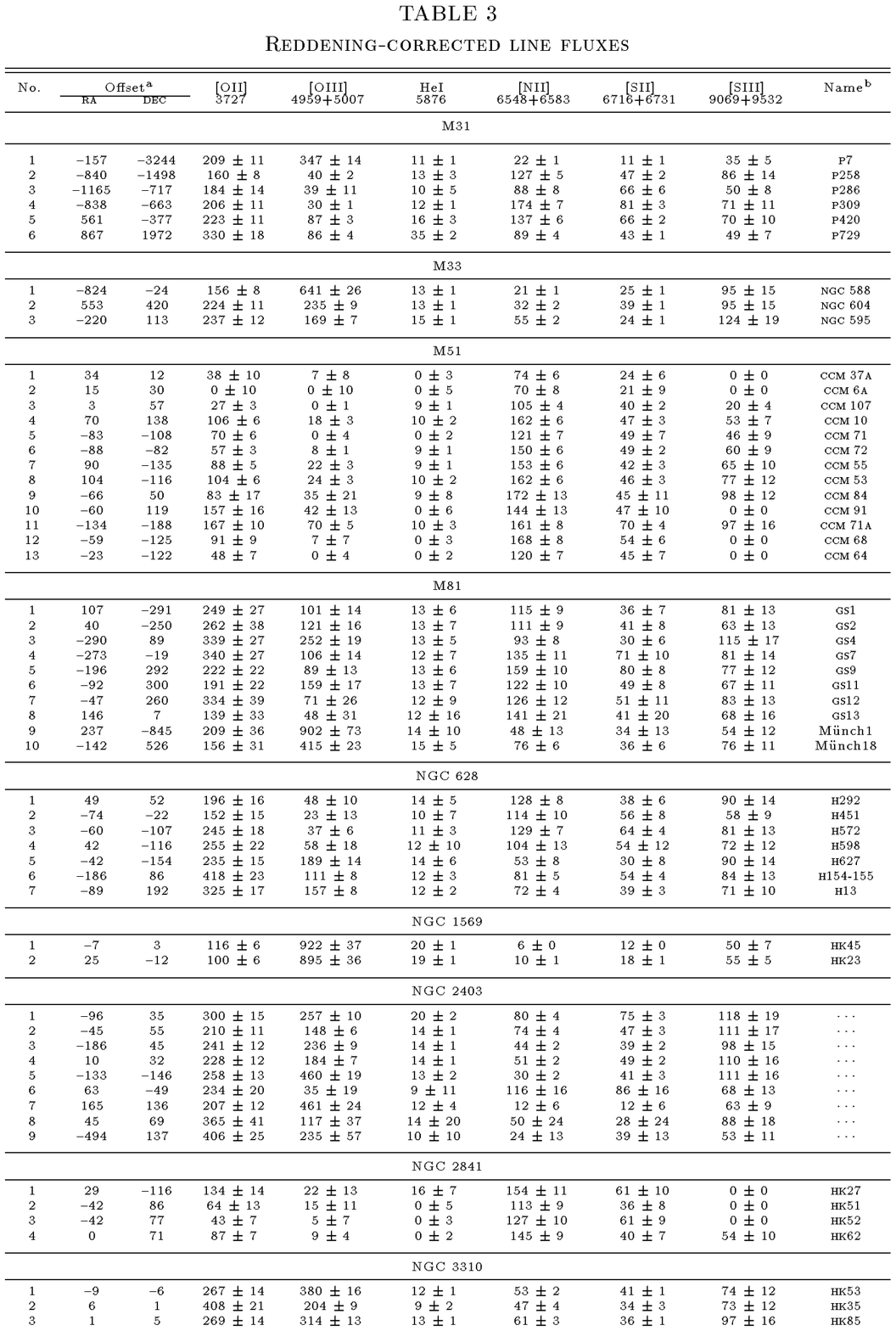}{23cm}{0}{100}{100}{-310}{-110}\label{galaxies.data}
\end{figure}

The emission line fluxes for the atomic transitions of interest were
measured with the IRAF\footnote[3]{IRAF is distributed by the National
Optical Astronomical Observatories, which are operated by AURA, Inc.,
under contract to the NSF.} task SPLOT, by integrating over the line
profiles. We assumed a stellar underlying continuum Balmer absorption
equivalent width of 2\angs, the mean value found by McCall \etal
(1985) in their \hii\/ region sample, and calculated the reddening
with the \halpha/\hbeta\/ and
\hbeta/H$\gamma$ ratios, assuming Case B recombination ratios for
an electron temperature of 10,000 K and density of 100 cm$^{-3}$
(Osterbrock 1989).  Table~\ref{galaxies.data} lists the measured
fluxes, corrected for reddening, for the most important emission
lines, as well as offsets in arcseconds from the galaxy nucleus.
Uncertainties in line fluxes were derived following the procedure
described in Kennicutt \& Garnett (1996), who analyzed
\hii\/ region spectra in M101 taken with the same setup as the
observations discussed in this work. The quoted errors were calculated
by adding in quadrature the contributions from the statistical noise,
the calibration ($\pm3$\%), the flat fielding ($\pm1$\%), the relative
scaling of the echellette orders ($\pm4$\%) and the uncertainty in the
reddening corrections.  A 20\% uncertainty was assigned to the
combined \siii\/ flux, as described earlier.

We performed a few tests in order to check the quality of the data.
As mentioned above the observed values of the
[SIII]\line9532/[SIII]\line9069 ratio are consistent with the
theoretical expectations. The density-sensitive
[SII]\line6716/[SII]\line6731 ratios fall near or below the
low-density limit (=~1.42), with an average value of 1.37 $\pm$ 0.02,
therefore justifying the assumption of low density (10--100 cm$^{-3}$)
made for the model nebulae (\cf
\S~\ref{models}). Finally, in Figure~\ref{mrs} our data (filled
circles) are compared to the data of McCall
\etal (1985, open circles) for two excitation sequences, 
[OII]/\hbeta\/ vs [OIII]/\hbeta\/ and [OIII]/\hbeta\/ vs
[NII]/\halpha.  As can be seen, the sequences for the two datasets
closely follow each other.

\section{PHOTOIONIZATION MODELS}\label{models}

In order to interpret the emission-line data for our \hii\/ regions, a
set of nebular models was calculated with the CLOUDY program (Ferland
\etal 1998; Ferland 1996), version 90.01. Details about the models and
more extensive results can be found in Bresolin (1997).

Given the spectral energy distribution (SED) of the ionizing source,
the ionizing photon luminosity, the chemical composition and the
geometry of the gas, the code solves for the ionization and thermal
equilibrium of the nebula. A relevant aspect of the version of the
code used for these calculations is the inclusion of Opacity Project
data (Seaton \etal 1994). For our models the important change
(approaching a factor of two) with respect to previous versions of the
code regards the O$^+$ photoionization cross section, which no longer
shows a jump at the 2s~--~2p edge. A further change in the
O$^{+}$~-~O$^{++}$ balance is caused by new collision strengths for
some O$^{++}$ lines.  To assess the effect of the new atomic data on
the final results, a set of single-star photoionization models
(described in the next section) were calculated with an older version
of CLOUDY ($v.\;84.12$), and compared with the results of
$v.\;90.01$. The newer version predicts an average decrease of the
[OII] line intensity of approximately 20\%, mostly counterbalanced by
an increase in the [OIII] emission.  The corresponding effect on the
empirical abundance indicator \x\/ translates into an approximate
0.025 dex increase in the (O/H) abundance scale, based on the
calibration given by Zaritsky \etal (1994).

For the remainder of this section the input parameters of the models
will be described; the relevant model results will be illustrated in
\S~\ref{generalresults}

\subsection{Input Parameters}
\subsubsection{Spectral Energy Distributions}
As mentioned above, two different SEDs were considered as the ionizing
source of the nebulae. The single-star models are simpler cases in
principle, as they can be interpreted in terms of one parameter,
\tstar, which is a function of the hardness of the output ionizing
spectrum. On the other hand, cluster models contain information on the
mass function and the temporal evolution of the ionizing clusters.

For single-star models, the line-blanketed LTE Kurucz (1992) stellar
atmospheres, which include the effects of bound-free opacity from
heavy elements, were adopted for consistency with the cluster
models. The $\log g=5$ atmospheres were chosen, in order to circumvent
problems of non-convergence of the hydrostatic LTE atmospheres at low
stellar surface gravity.  The metallicity was varied in accordance
with the gas abundance. Nebular models were calculated at
Z~=~0.25~\zsun, \zsun\/ and 2~\zsun, and for
\teff\/~=~37,500 K, 40,000 K, 45,000 K, 50,000 K and 60,000 K.  The
sensitivity of the results to the choice of stellar atmosphere models
will be discussed later.

An alternative approach is to calculate SEDs for star clusters with
different IMFs. This method was applied by McGaugh (1991) for zero-age
clusters, and taking into account the temporal evolution of the
clusters by Garc\'{\i}a-Vargas \& D\'{\i}az (1994), Cervi\~{n}o \&
Mas-Hesse (1994), Garc\'{\i}a-Vargas \etal (1995), and Leitherer \&
Heckman (1995).

We adopted the SEDs calculated by Leitherer \& Heckman (1995), as
presented in Leitherer \etal (1996a). These SEDs were also calculated
for different metallicities (those at 0.25 \zsun, \zsun\/ and 2
\zsun\/ are relevant to the present work) and IMFs ($\alpha~=~2.35$ and M$_{up}$~=~100 \msun,  
$\alpha~=~2.35$ and M$_{up}$~=~30 \msun, $\alpha~=~3.35$ and
M$_{up}$~=~100 \msun, where the IMF is given by
$dN/dm~\propto~m^{-\alpha}$ between M$_{low}=1$ \msun\/ and M$_{up}$),
at 1 Myr intervals from 1 to 25 Myr.  The stellar models of Maeder
(1990) and Maeder \& Meynet (1988) were used for the synthesis,
together with Kurucz's (1992) atmospheres, supplemented by the non-LTE
models of Schmutz, Leitherer \& Gruenwald (1992) for stars with strong winds. We used
the results of both the instantaneous burst case, in which star
formation occurs over a timescale short compared to the age of the
model, and the continuous star formation case (star formation constant
with time). We considered ages from 1 to 6 Myr, since at later times
most clusters would lack the necessary ionizing luminosity to produce
an \hii\/ region (Garc\'{\i}a-Vargas \& D\'{\i}az 1994).  For further
details on the cluster SEDs, the reader is referred to the original
papers.

\subsubsection{Nebular Gas Parameters}
The abundance of the gas has been made consistent with the metallicity
of the ionizing stars. The solar elemental abundances are adopted as
defined in Grevesse \& Anders (1989). To take into account depletion
of refractory elements onto dust grains, the elements Mg, Al, Ca, Fe,
Ni and Na were depleted by a factor of 10, and Si by a factor of two,
relative to the solar abundance (consistent with Garnett \etal 1995,
Garnett \& Kennicutt 1998).  The He abundance was scaled according to
Y~=~Y$_p$~+~($\Delta$Y/$\Delta$Z)\thinspace Z, where Y $=4y/(1+4y)$,
$y$~=~He/H by number, and Z is the metallicity (\zsun~=~0.02). We have
adopted Y$_p=0.23$ (Pagel \etal 1992), and $\Delta$Y/$\Delta$Z=2.5.
Table~\ref{elements.table} summarizes the adopted solar composition,
with the above-mentioned depletions taken into account.

The model geometry is spherical, with a constant gas density
n~=~50~cm$^{-3}$, and with a filling factor $\epsilon$ adjusted so as
to obtain models with three different ionization parameters, $\log U =
-2, -3$ and $-4$. The dimensionless ionization parameter $U$ is
defined as
\begin{equation}
U = \frac{\qh}{4\pi R_s^2 nc}
\end{equation}

\noindent
where \qh\/ is the number of hydrogen-ionizing photons (E~$>$~13.6 eV)
emitted per second, R$_s$ is the Str\"{o}mgren radius, n is the
hydrogen density and c is the speed of light. $U$ depends on the
electron temperature, T$_e$, through the Str\"{o}mgren radius
dependence on the case B recombination coefficient $\alpha_B(T)$:
\begin{equation}
R_s = \left(\frac{3}{4\pi}\frac{\qh}{\alpha_B n^2
\epsilon}\right)^{1/3}
\end{equation}
with $\alpha_B(T)\propto T_e^{-1}$,
so that we can rewrite
\begin{equation}
U = A(\qh n\epsilon^2)^{1/3}
\end{equation}
with $A\propto T_e^{-2/3}$. We had therefore to iterate a few times, from
an initial guess for the filling factor, to obtain the desired $\log
U$.  A few models were calculated for n~=~10~cm$^{-3}$ and
n~=~150~cm$^{-3}$, to estimate gas density effects, which are quite
strong at higher metallicities (D\'{\i}az \etal 1991).

We fixed the total number of ionizing photons to \qh\/~=~10$^{51}$
s$^{-1}$, the equivalent of $\sim$90 O7 V stars (Vacca 1994), and
representative of bright extragalactic \hii\/ regions. The exact value
of \qh\/ is of no practical importance, since it acts as a scaling
factor, and models with a given ionizing spectrum shape and differing
\qh, n and $\epsilon$, but with the same $U$ are homologous, producing
the same emission-line relative strengths (as long as the density is
below the critical limit for collisional de-excitation) and the same
nebular ionization structure. The corresponding initial cluster masses
are large enough ($>1\times10^4$~\msun) that our models remain
unaffected by the stochastic effects described by Garc\'{\i}a-Vargas
\& D\'{\i}az (1994) and Cervi\'{n}o \& Mas-Hesse (1994).  We also
emphasize that CLOUDY computes nebular models in which a compact
ionizing source lies at the center of a spherically symmetric gas
distribution. We do not expect significant changes in the case of a
more extended ionizing source (\eg\/ a loose OB association) as long
as it remains unresolved by the spectrograph slit, since the
emission-line properties are controlled by a quantity integrated over
the volume of the nebula (the ionization parameter).

All of our models are ionization-bounded, with the calculation stopped
at a gas temperature of 100 K (the optical opacity of the gas is
negligible even at a much higher temperature). The models include a 30
km s$^{-1}$ turbulent velocity term, consistent with observed
linewidths of giant \hii\/ regions (Arsenault \& Roy 1988). This
additional velocity field increases the line widths (already broadened
by thermal motions), affecting the energy budget of the \hii\/
regions. A cooling of the nebulae results through infrared
fine-structure lines, which can become optically thin as a result of
the line broadening.  The effects of dust have not been taken into
account, except for the depletion of refractory elements.

For the interested readers, the results of our models can be
retrieved, in the form of tables and diagnostic diagrams, by anonymous
ftp, by contacting the first author.

\section{RESULTS}\label{generalresults}

A useful quantity that can be readily estimated from the stellar
atmosphere SEDs is the ratio of He$^0$ ionizing photons ($\lambda\leq
504$\angs) to H ionizing photons ($\lambda\leq 912$\angs),
$Q_{He^0}$/\qh, for each metallicity and \teff.  For a particular
atmosphere model, this ratio describes the hardness of the radiation
field, and has been used to define an equivalent ionizing temperature,
T$_{eq}$, as the \teff\/ of a star whose SED produces a given
$Q_{He^0}$/\qh\/ (Mas-Hesse \& Kunth 1991, Garc\'{\i}a-Vargas
\& D\'{\i}az 1994). A calibration of this ratio for the Kurucz (1992)
$\log g=5$ atmospheres is given in Table~\ref{qratio.table}. Clearly
the radiation field becomes harder as \teff\/ increases, and with
lowering metallicity at a fixed temperature. The sensitivity to
abundance is largest at high temperatures, becoming progressively
smaller at lower \teff.  The abundance effects can be quite important,
since they can be equivalent to a change in \tstar\/ of as much as
10,000~K between 2~\zsun\/ and 0.1~\zsun.

The $Q_{He^0}$/\qh\/ ratios for the cluster models are shown in
Table~\ref{qratioclusters.table}. As pointed out by Garc\'{\i}a-Vargas
\etal (1995), the evolution of T$_{eq}$ is non-monotonic at higher
metallicities, due to the hardening of the radiation after the
predicted W-R phase sets in. This effect disrupts the
T$_{eq}$~--~metallicity relationship valid for ages smaller than about
3 Myr, for which T$_{eq}$ drops by 5000-6000~K (from 50,000 to
45,000~K at t = 1 Myr) between 0.1~\zsun\/ and 2~\zsun. The presence
of the W-R stars strongly affects the results 2-3 Myr after the burst
of star formation.  The effect is stronger with increasing
metallicities, because of the larger number of W-R stars produced.

A direct comparison of our model results with those of 
other authors is made difficult by the use of different input
parameters and photoionization codes. However, a small set of solar
composition models, with n~=~10~cm$^{-3}$, was compared to the
models by Stasi\'{n}ska \& Leitherer (1996). The latter were
calculated with the code PHOTO (Stasi\'{n}ska 1990) and the same
cluster SEDs used in the present work, for different cluster
masses. Due to the evolution of the massive stars, the total ionizing
luminosity (and hence the ionization parameter) is a function of
cluster age, while in our models we have kept it constant, in order to
set the ionization parameter to the desired values. The Stasi\'{n}ska
\& Leitherer models for cluster masses M$_*=10^3$ and 10$^6$
\msun\/ and ages between 1 and 6 Myr (M$_{up}$~=~100 \msun) have an
ionization parameter in the range --1.6 to --2.9, and can therefore be
compared to our $\log U\,=\,-2$ and --3 models. Figure~\ref{comp}
shows the comparison regarding [OIII]\line5007 vs \hbeta\/ and
[OII]\line3727 vs \hbeta\/. There is a good agreement, considering the
differences in ionization parameter between the two sets of models,
and the different photoionization codes used.

\subsection{Ionization Parameter}\label{ionpar}
The model results indicate that the ionization parameters adopted for
the models well bracket our data, which are clustered around $\log
U=-3$. This is demonstrated in Figure~\ref{siisiii_r23k}, which plots
[SII]\lines6717,6731/[SIII]\lines9069,9532 vs the abundance parameter
\x\/ for our sample. D\'{\i}az \etal (1991) gave a calibration of
$U$ in terms of this ratio, which has been refined by Garnett \&
Kennicutt (1998) to include the small metallicity dependence.
Similarly we have used a surface fit to our models in order to derive
the following relation:

\begin{equation}
\log U = (-3.35 \pm 0.04) - (1.54 \pm 0.05) \log([SII]/[SIII]) - (0.08
\pm 0.06) \log(R_{23})\label{eqtn}
\end{equation}

According to this calibration and the observed [SII]/[SIII] ratios the
ionization parameter in our sample is in the range $\log U = -3 \pm
0.6$.  The comparison of the models with the observations suggests
also that for the most metal-poor objects ($\sim$ 0.2~\zsun) the
ionization parameter is about 4 times larger than for \hii\/ regions
around the solar metallicity. This finding is in agreement, at least
qualitatively, with previous investigations of \hii\/ galaxies
(Campbell 1988, Stasi\'{n}ska \& Leitherer 1996).

\subsection{The N/O abundance ratio}\label{additional}
The [NII]/[OII] model sequences do not reproduce the observations at
low abundances when a constant N/O ratio is adopted. The actual data
show weaker [NII] lines relative to [OII] when compared to the
theoretical values.  Theory and observations could be reconciled by
considering a metallicity-dependent N/O ratio. Evidence for such an
effect in spirals has already been shown by several authors (\eg\/
McCall \etal 1985, Fierro, Torres-Peimbert \& Peimbert 1986, D\'{\i}az
\etal 1991, Garnett \& Kennicutt 1998), but there is no general
agreement on the functional relation between N/O and O/H, if it exists
at all (different spirals having possibly different N/O gradients,
Henry \& Howard 1995; these authors, however, neglected primary
nitrogen in deriving their gradients).  In low-abundance irregular
galaxies the N/O ratio is found to be independent of O/H (Garnett
1990), supporting the idea of a primary origin for nitrogen at low
metallicity: nitrogen is produced out of C and O synthesized during
the evolution of stars. An abundance-sensitive N/O ratio would imply
the presence of a secondary component, that is N produced by C and O
already present in stars at their birth. The situation is less clear
for S/O, for which the existence of a O/H dependence is uncertain (for
example, see contradicting results in D\'{\i}az \etal 1991 and Garnett
1989).  To represent a secondary component superposed on a primary
component for N in spiral galaxies, a few models were calculated
assuming the following relations (Garnett \& Kennicutt 1998):
\begin{equation}
\eqnum{5a}
\log N/O=-1.5+(\log O/H+3.7)\qquad(\log O/H > -3.7)
\end{equation}

\begin{equation}
\eqnum{5b}
\log N/O=-1.5\phantom{+(\log O/H+3.71)}\qquad(\log O/H < -3.7)
\end{equation}
\setcounter{equation}{5}

The results (Fig.~\ref{no}) are in much better agreement with the
observations, lending more credibility to the presence of a secondary
component for nitrogen in spiral galaxies.  The models can now
reproduce the steepening of the [NII]/[OII] vs
\x\/ relation that occurs in the data at Log~\x~$>~$0.5. 
The main secondary effect of the abundance-dependent N/O ratio is a
shift of the 2 \zsun\/ models to lower \x\/ values, due to the
increased cooling of the nebulae caused by the larger abundance
fraction of nitrogen. Models with a fixed, solar N/O ratio would
therefore overestimate the metallicity for high-abundance objects.
The models presented in this paper were calculated with a constant N/O
ratio; however the effects of a variable ratio for all the diagnostics
but those involving [NII] are, for all our purposes, marginal and
limited to Z~$>$~Z$_\odot$.

An important point to make about the [NII]/[OII] sequence is that it
refers to 20 spiral galaxies of different Hubble type. The tightness
of the sequence implies that the relation between N/O and O/H is
similar for all these galaxies, whatever the exact functional form
might be. While for a few individual galaxies (\eg\/ M33, see
V\'{\i}lchez \etal 1988, or M81, see Garnett \& Shields 1987) a N/O
gradient seems not to be required, a `universal' abundance-dependent
N/O ratio seems to be necessary to explain the bulk of our
observations.

\subsection{The \hbeta\/ Equivalent Width}\label{ewresults1}
Bresolin \& Kennicutt (1997) studied the distribution of the \halpha\/
equivalent width [EW(\halpha)] of \hii\/ regions in ten galaxies of
different Hubble type to test the idea that different upper mass
limits of the IMF could be responsible for the changes in star forming
region properties as a function of morphological type. The
photometrically-determined EW(\halpha) values showed no dependence on
Hubble type.  The distribution of the equivalent width of \halpha\/
and \hbeta\/ for our \hii\/ region sample confirms this result. Since
the equivalent width of the Balmer emission lines is a measure of the
ratio of the number of hot, massive stars (producing the emission line
flux through ionization of the nebula) to the total number of stars
(producing the continuum), our result points toward the invariance of
the IMF across the morphological sequence.

The data show a dependence of \ew\/ on metallicity, with a large
scatter, likely due to an age spread in the \hii\/ region sample. A
similar behavior was already measured by Searle (1971) on a small
sample of \hii\/ regions in M101. This result was interpreted by
Shields \& Tinsley (1976) as evidence for a gradient in the
temperature of the hottest exciting stars, which in turn led them to
hypothesize an abundance-dependent upper mass limit for star
formation.  The photoionization models, however, fail to reproduce the
observed sequence of \ew\/ vs \x. It is a well-known fact that the
observed \ew\/ values of extragalactic \hii\/ regions are much lower than the
theoretical predictions. The usual interpretation is that an
underlying population of stars older than the ones responsible for the
ionization of the nebulae is contributing to the stellar continuum,
thus diluting the ionizing radiation and reducing the equivalent width
(McCall \etal 1985). Diaz
\etal (1991) proposed the coexistence of stellar clusters of different
ages within giant \hii\/ regions, as suggested by spatially resolved
observations (Skillman 1985, D\'{\i}az \etal 1987). Internal
extinction due to dust which would affect the nebular lines, but not
the continuum, has been proposed as an alternate solution (Mayya \&
Prabhu 1996, Garc\'{\i}a-Vargas \etal 1997).  The dependence of \ew\/
on both cluster age and upper IMF, combined with the problem just
described, make it difficult to derive a \teff--Z relationship from
\ew. Therefore we will not attempt to use the observed \ew\/
to infer a \tstar\/ gradient.

\subsection{Cluster Ages}
A \tstar-metallicity relation is expected for young clusters (1-2 Myr
after initial burst), as a result of metallicity-dependent effects on
the stellar output flux. The onset of the W-R phase for massive stars
which occurs after 2-3 Myr of the initial burst would disrupt this
correlation for a mixed-age sample of clusters, as shown by
Garc\'{\i}a-Vargas \etal (1995).  The observation of a \tstar\/
gradient in our sample of \hii\/ regions (\S~5) would therefore imply,
if theory and models are correct, that an age limit
exists. Interestingly, this would agree with the fact that the
instantaneous burst cluster models are only able to reproduce the data
for young cluster ages (t~$<$~3 Myr). The theoretical predictions
diverge considerably from the observations at later times. This does
not exclude the likely presence of older
\hii\/ regions in our sample, but the average of the \hii\/
region sequences is best reproduced by the 1 and 2 Myr models, when we
make the assumption that the ionizing sources are {\em single
population} clusters. As an example, we show the diagnostic diagrams
which summarize the model results for [OII]/[OIII] vs \x\/ in
Figure~\ref{oiioiii_r23b}.  Here the data (solid dots) are compared
with the models, calculated for three different values of the
ionization parameter and for three different cluster IMFs, as
explained in \S~\ref{models}. It can be seen that the hardening of the
stellar radiation predicted after 2 Myr from the initial burst
produces line ratios which cannot in general be reconciled with the
observations (see also Fig.~\ref{eta_r23b}).  There remains the
possibility that the nebular models are at fault, even though
state-of-the-art codes and models have been adopted for both the
gaseous and the ionizing source treatment.  Alternatively some of the
predictions of the stellar evolutionary models could be incorrect.  In
short, our results imply that either the current evolution models
overpredict the hardness of the emitted Lyman continuum spectrum, or
that some physical mechanism, such as disruption by stellar winds
and/or supernova explosions, prevents us from observing \hii\/ regions
older than a few million years.

The continuous star formation cluster models often provide as good a
fit to the observations as the single-age burst models. Although most
published photoionization models are based on instantaneous burst
models, this suggests that continuous models should also be considered
as valid alternatives. Star formation extending over time (an age
spread of a few Myr), as in the case of individual star clusters
formed in subsequent bursts, but close enough to remain unresolved by
the spectrograph slit, is a scenario that deserves further
investigation.  This is the case of 30 Dor, whose central ionizing
cluster has an age of $\sim$3 Myr; there is however evidence for
previous episodes of star formation, as indicated by the presence of
supergiants (Walborn 1991). These components of different age are
spatially resolved, but they would not be if 30 Dor were at a much
larger distance.  Recent work (Garc\'{\i}a-Vargas \etal 1997,
Gonz\'{a}les-Delgado \etal 1997) indicates the coexistence of young
($\sim$3 Myr) and old ($\sim$8 Myr) components in individual
starbursts, supporting the idea of an age spread, and suggesting star
formation in multiple, subsequent bursts as a likely mechanism in
giant \hii\/ regions.

\section{TEMPERATURE OF THE IONIZING STARS}\label{teff}
Shields \& Searle (1978) were the first to point out the importance of
the [SIII]\lines9069,9532 lines to constrain the ionization structure
of \hii\/ regions in the absence of temperature-sensitive lines, since
sulfur is an effective coolant through these near-IR forbidden lines
and the fine-structure lines at 19 and 33 $\mu$m.  Mathis (1982, 1985)
later introduced a method to determine the relative effective
temperature scale, based on \sp/\spp\/ and \op/O. The use of ratios of
ionic stages of the same element makes this method almost independent
of nebular abundance. However, absolute values of \tstar\/ are much
more difficult to obtain, due to the dependence on the stellar
atmospheres adopted for the nebular models. Besides, there is a
saturation effect for very hot stars, so that the method becomes less
useful at high temperatures (Garnett 1989).

A `radiation softness' parameter $\eta$ was introduced by V\'{\i}lchez
\& Pagel (1988) as a modification of Mathis' procedure, and defined
as
\begin{equation}
\eta = \frac{\op/\opp}{\sp/\spp}.
\end{equation}
It is almost independent of ionization parameter, reddening, electron
density and temperature. The observed quantity is given by
\begin{equation}
\eta\prime =
\frac{\rm [OII]\lines3726,3729\;/\;\rm[OIII]\lines4959,5007} {\rm
[SII]\lines6717,6731\;/\;\rm [SIII]\lines9069,9532.}
\end{equation}

As a measure of the hardness of the radiation field, this parameter
provides us with an indicator of the \teff\/ of the exciting stars of
\hii\/ regions.  The large difference in the ionization potentials of
\op\/ (35.1 eV) and \sp\/ (23.2 eV) is responsible for the sensitivity
of $\eta$ to the spectral energy distribution of the ionizing
radiation.  Spatially resolved observations of the Orion Nebula,
NGC~604 in M33 and 30 Doradus in the LMC confirm the reliability of
this method (V\'{\i}lchez \& Pagel 1988, Mathis \& Rosa 1991),
although Ali \etal (1991) questioned it on the basis of possible
effects of dielectronic recombination on the sulfur lines (the
relative coefficient has not been computed yet).  Applications of the
$\eta$ method to study the physical properties of extragalactic \hii\/
regions include the works of D\'{\i}az \etal (1987), V\'{\i}lchez \&
Pagel (1988), D\'{\i}az \etal (1991) on the high-metallicity \hii\/
regions in M51, and Garnett \& Kennicutt (1998) on a large sample of
objects in M101.

There is a general agreement about the existence of a metallicity
dependence of the hardness of the radiation for \hii\/ region ionizing
clusters. For example, in a sample of high-excitation
\hii\/ regions Stasi\'{n}ska (1980) found that \teff\/
increases statistically with decreasing abundance. The same conclusion
has been reached by studies of the generally metal-poor
\hii\/ galaxies (Campbell
\etal 1986, Melnick 1992, Cervi\~{n}o \& Mas-Hesse 1994).
V\'{\i}lchez \& Pagel (1988) confirmed this trend based on a
compilation of \hii\/ region data from several authors.  At any given
abundance the spread in \teff\/ typically amounts to $\sim$5,000~K,
which can be due to several combined causes, such as evolutionary
effects, uncertainties in the models, and differences intrinsic to the
ionizing population.  The methods used in these works were based on
the determination of electron temperatures, mainly via the
[OIII]\line4363 emission line, and are henceforth limited to low
abundance (generally Z~$<$~0.4~\zsun).

Different authors have instead debated over the cause of the \tstar\/
gradient, whether it is a reflection of a change in the upper IMF
(\eg\/ Campbell \etal 1986), or simply an effect of stellar evolution
at different metallicities (McGaugh 1991). To illustrate the latter,
we show in Fig.~\ref{geneva} the metallicity--\teff\/ relation
predicted by the Geneva stellar evolutionary models for different
stellar masses, and for ages of zero and one Myr (Schaerer \etal
1993b, and references therein). For massive stars a decrease of the
order of 5,000~K or more, up to $\sim$10,000~K, exists in the
metallicity range 0.05~\zsun\/ -- 2~\zsun. The effect is stronger at
low abundance, where the gradient is steeper, which seems to be in
agreement with \hii\/ region observations.

In the following we examine our data in search for variations of
\tstar\/ with abundance, by means of the analysis of the $\eta\prime$
parameter. Since our sample covers the range in metallicity from about
0.2~\zsun\/ to above solar, we are able to extend previous work in
order to check if the \tstar\/ trend continues at higher
abundances. However, as mentioned above, the expected change in this
metallicity range is rather limited, making it more difficult to
detect. The fact that at high abundance the modelling of \hii\/
regions becomes problematic (due also to the lack of electron
temperatures, forcing us to use an empirical calibration for the
metallicity) complicates the interpretation of the data. In addition,
the dependence of the results on the stellar atmospheres adopted, and
the sensitivity to the ionization parameter make the precise \tstar\/
scale, and in part its gradient, model dependent.

\subsection{\tstar\/ gradient from single-star models}
Figure~\ref{eta_r23k} shows the $\eta\prime$ vs \x\/ diagnostic
diagram for our data and single-star models, and illustrates the main
difficulties in deriving a \tstar\/ gradient from $\eta\prime$. The
main complications are its dependence on \x\/ (at constant T and
$U$) and the weaker sensitivity to \tstar\/ in the high-\tstar\/
range. Changes in the ionization parameter also have considerable
effects on the results.  We finally remind the reader about the rather
mild \tstar\/ dependence on metallicity which is expected when the
abundance is larger than about half solar.  The \hii\/ region parent
galaxies are labeled in the $\eta\prime$~--~\x\/ plot of
Figure~\ref{labelgalaxy}, where the sample has been divided into
early-type (Sa--Sb; upper panel) and late-type (Sbc--Sm; lower panel)
spirals. We also include spatially resolved observations of the 30
Doradus nebula in the LMC by Kennicutt \& French (Kennicutt {\it et al.}, in
preparation), which show that the dispersion in line ratios for a
typical supergiant \hii\/ region is roughly of the same magnitude as
the dispersion in the whole sample.  When compared with the models of
figure~\ref{eta_r23k}, this leads us to conclude that accurate
\tstar\/ determinations cannot be obtained through $\eta\prime$. The
strength of the method lies in the possibility of detecting
large-scale trends of \tstar\/ in a large sample of high-abundance
\hii\/ regions.

In order to simplify the interpretation of the data, in
Fig.~\ref{logn.4}a we have divided the \hii\/ region data into two
ionization parameter classes ($\log U$ above and below $-2.75$),
according to the measured [SII]/[SIII] ratio and equation
(\ref{eqtn}), and superposed the models calculated for $\log U = -2.5$
and $-3.0$ at various stellar effective temperatures for the Kurucz
atmospheres.  The comparison of the models with the observed sequence
suggests a metallicity dependence of \tstar, in the sense that the
upper bound of \tstar\/ shifts to lower temperatures at higher
abundances, in a fashion similar to what Campbell (1988) noted for
metal poor \hii\/ galaxies.  Within our sample this effect is more
easily seen in the high ionization parameter range ($\log U >
-2.75$). Quantifying it is however much more difficult, and, as the
bottom part of Fig.~\ref{logn.4}a shows, the models do not agree with
all the observational data.  This discrepancy was seen by, among
others, Skillman (1989) and Garnett (1989), who pointed out that the
S$^+$/S$^{++}$ ratio is underpredicted (or O$^+$/O$^{++}$
overpredicted) by nebular photoionization models.  It has been
suggested that the origin of the disagreement lies in the sulphur
atomic parameters (Garnett 1989) or in density inhomogeneities of the
nebular gas (Rosa 1993).  Garnett (1989), after examining the effects
of different stellar atmospheres on the model predictions for these
two ionic ratios, concluded that only relative changes of \tstar\/ can
be inferred from $\eta$.  Before reaching any conclusion on the
existence of a \tstar\/ gradient at high abundance we will also
briefly consider the effects of changing the input stellar
atmospheres, including the most recent, state-of-the-art hot star
atmosphere models.

\subsection{Effects of input stellar atmospheres on the interpretation
of the \tstar\/ gradient}

The choice of stellar atmospheres, such as the widely used models by
Mihalas (1972) and Kurucz (1979, 1992), is responsible for well-known
differences in the predicted emission line ratios when computing
\hii\/ region photoionization models (Skillman 1989, Evans
1991). These models have well-recognized inadequacies, namely the
absence of line blanketing (Mihalas) or treatment of plane-parallel
atmospheres in LTE (Kurucz). The inadequacy of these models for
early-type stars derives from severe departures from LTE and the onset
of radiation-driven stellar winds (Kudritzki \& Hummer 1990, Kudritzki
1998). Enormous progress has been made in the last decade in stellar
atmosphere modeling, in being able to treat theoretically the effects
of departures from LTE, opacity from metal lines, stellar winds, and
spherical extension on the computed hot star spectral energy
distribution and ionizing flux (Gabler \etal 1989, Najarro \etal 1996,
Schaerer \etal 1996a,b). The modifications with respect to previous
calculations are profound. For instance, the predicted flatter
ionizing flux distribution in the He~I continuum (below 504~\AA) has
already offered a solution for a long-lasting discrepancy between
observations and nebular models regarding the intensity of the [NeIII]
lines, severely underpredicted (up to a factor of 10 or more) when
adopting the Kurucz atmospheres (Sellmaier
\etal 1996, Stasi\'{n}ska \& Schaerer 1997). 
Our understanding of the stellar populations responsible for the
ionization of extragalactic giant \hii\/ regions is also affected, as
shown by the possible explanation of high-excitation nebulae in terms
of `normal' massive O stars, instead of more exotic ionizing sources
(Gabler \etal 1992).

Various predictions of the spectral energy distribution of hot stars
are becoming available (Pauldrach \etal 1994, Sellmaier \etal 1996,
Najarro \etal 1996, Schaerer \& de Koter 1997, Pauldrach \etal
1998). While basically the same physical processes are taken into
account, considerable differences exist in modeling techniques, for
example in computing the occupation numbers for the different ions, so
that different \hii\/ regions ionization structures would be
predicted. While a detailed comparison is out of the scope of the
present paper, we have computed nebular models based on stellar
atmospheres different from Kurucz's, in order to verify that the
general, qualitative conclusions remain unchanged. We have therefore
computed nebular models using the most recent early-type star
atmospheres available: the CoStar models (Schaerer \& de Koter 1997),
available for two different metallicities (\zsun\/ and 0.2~\zsun), and
recent hydrodynamical non-LTE calculations of expanding atmospheres
kindly made available to us by the Munich group (see Pauldrach \etal
1998).  The latter, too, were calculated for Z~=~\zsun, 0.2~\zsun. The
2~\zsun\/ nebular models were computed with solar abundance stellar
atmospheres.  Additional models were computed with the Mihalas (1972)
atmospheres, keeping the abundance of the stellar ionizing source
fixed. All the remaining parameters were the same as described earlier
for the Kurucz atmospheres.  

Without discussing the different theoretical approaches, for which we
refer the reader to the above-mentioned papers, or the apparent merits
and faults of the atmosphere models used in this comparison, we
present in Figs.~\ref{logn.4}b--\ref{logn.4}d the resulting
models in the $\eta\prime$~--~\x\/ diagram, with the same partition
of the data into two ionization parameter bins as in
Fig.~\ref{logn.4}a.  The figures demonstrate very clearly that the
absolute \tstar\/ scale greatly depends on the adopted input ionizing
flux. It is particularly interesting that the two most recent non-LTE
expanding atmosphere models give very different results in the lower
\teff\/ regime; future work on \hii\/ region modeling can in principle
be very useful in solving these discrepancies. Concentrating however
on the points in common, we find that all models are consistent with
higher \tstar\/ values ($\geq50,000$~K) at the lowest abundances,
while at high abundance \tstar\/ levels off at a smaller value,
between 35,000~K and 45,000~K, depending on the model atmospheres.  We
conclude that the data cannot be interpreted with a constant effective
temperature at lower abundance: somewhere between
\zsun\/ and 0.2~\zsun\/ (we cannot specify a precise value with the present
models and observations) the average $\eta\prime$--\x\/ relation
steepens more than the constant temperature models. Constant \tstar\/
cannot be ruled out for higher abundances.

\subsection{Trends at low and high abundance}\
The line ratios studied earlier seem to conspire against a clear
definition of a temperature dependence on metallicity. But, as seen at
the beginning of this section, direct electron temperature
determinations in nebular spectra below Z~$\simeq$~0.4~\zsun\/ have
led to the notion of a \teff\/ gradient for metal poor objects. How
can this be reconciled with the present work?  In the diagnostic
diagram of Figure~\ref{dennefeld} (top) the open stars mark the
position of a few galactic and Magellanic Cloud \hii\/ regions with
metallicity 0.2~$<$~Z/\zsun~$<$~0.5 from the work of Dennefeld \&
Stasi\'{n}ska (1983). These are rather low luminosity \hii\/ regions
when compared with the objects in our sample, but they might serve for
the purpose of better understanding the diagnostic diagrams.  The
models shown for comparison were computed with the Munich group
atmospheres at $\log U = -2.0$, since these are high ionization
parameter objects. No metallicity effect on
\tstar\/ can be derived from this plot. For comparison the bottom part of
Figure~\ref{dennefeld} shows the objects in our \hii\/ region sample
for which an abundance based on electron temperature measurements is
available in the literature. Here the models refer to $\log U = -2.5$,
in agreement with the average $<$$\log U$$>$ $= -2.6$ for the data
points. For these two sets of \hii\/ regions we compare in
Fig.~\ref{oh} the observed $\eta\prime$~--~O/H relation with the one
predicted by nebular models (Munich group atmospheres, $\log U =
-2.5$). The 30 Dor position is marked by the open star. The
\tstar\/ trend with metallicity becomes now clearer, and well agrees with
previous work (see for example Fig.~6 in Stasi\'{n}ska 1980, or
Fig.~4.3.2 in Melnick 1992).

What can be said at the opposite (metal-rich) end of the abundance
range? As mentioned earlier, the $\eta\prime$ data can be consistent
with a constant \tstar\/ above the solar metallicity. This statement
can be checked by examining another important diagnostic at low
temperatures, the intensity of the neutral helium line He~I \line5876
with respect to the \hbeta\/ emission line intensity. This line ratio
provides a particularly sensitive diagnostic in the 35,000--40,000~K
range, and tends to saturate at higher temperatures; unlike
$\eta\prime$, it is only weakly dependent on ionization parameter. The
corresponding observations and nebular models are shown in
Fig.~\ref{hei}, which suggests that \hii\/ regions with above-solar
abundance define an upper bound in temperature at
\tstar~$\simeq35,000$~K.  Comparison of our models with the data
strongly suggests that \tstar\/ drops sharply below 40,000~K for
Z~$>$~\zsun.  We regard this result to be quantitatively more accurate
than the one obtained with $\eta\prime$, given the uncertainties
already described for the latter. Qualitatively, there is no
contradiction between the two results.

In conclusion, the single-star models support the idea of an important
dependence of \tstar\/ on metallicity over the whole range of
abundance displayed in our sample. However, we must use some caution
in the interpretation of this result, which is dependent on the
various assumptions made in our nebular models, besides the
uncertainties in the input ionizing fluxes.

\section{DISCUSSION: DOES THE IMF CHANGE WITH METALLICITY ?}\label{imf}
Investigations of possible variations of the IMF in star forming
regions have often been stimulated by the Shields and Tinsley (1976)
suggestion of a dependence of the upper mass limit on metallicity,
M$_{up}\propto Z^{-a/2}$, with $a\simeq 1$. This followed from
theoretical considerations by Kahn (1974), who argued that the
increase of the interstellar grain opacity with $Z$ would, as an
effect of the consequent increase in radiation pressure, prevent
further accretion onto a forming star, and set an upper limit to the
luminosity-to-mass ratio. We note, however, that Shields and Tinsley's
theoretical relation between \ew\/ and T$_u$ (temperature of the
hottest stars present) was, considering the measured gradient of
\ew\/ in M101, only marginally consistent with a dependence of
M$_{up}$ on $Z$.  Adopting a Salpeter (1955) slope ($\alpha=-2.35$)
for the IMF, instead of the steeper value used by Shields and Tinsley
($\alpha\simeq -4$), makes the inferred T$_u$ gradient inconsistent
with a significant dependence of M$_{up}$ on metallicity (Scalo 1986).

Direct counts of massive stars in resolved star clusters and OB
associations in our Galaxy and the Magellanic Clouds do not show a
variation of the IMF with the galactic environment or the metallicity,
at least for metallicities equal to solar or below (Massey \etal
1995a,b, Hunter \etal 1997).  For more distant star forming regions,
studies of the nebular recombination lines (\eg\/ Garc\'{\i}a-Vargas
\etal 1995, Stasi\'{n}ska \& Leitherer 1996) and of the UV resonance
lines (Robert, Leitherer \& Heckman 1993) have also pointed to a
`universality' of the IMF. In this work we have compared emission-line
diagnostics for a large \hii\/ region sample to theoretical
predictions. We now conclude with some comments on our results, and
their implications for constraints on the upper IMF variations.

Our cluster photoionization models often are difficult to interpret
unambiguously, due to the fact that different combinations of age,
ionization parameter and IMF can reproduce the observed sequences in
the diagnostic diagrams, as seen in Fig.~\ref{oiioiii_r23b}.
Ambiguities are also introduced by the sensitivity of the inferred
stellar temperatures to the choice of stellar atmosphere models.  A
further example is presented in Fig.~\ref{eta_r23b} for the
$\eta\prime$ parameter. The two IMFs with slope $\alpha=2.35$ and
$\alpha=3.30$, but having the same upper mass limit $100$~\msun, are
virtually indistinguishable from a comparison of the observations with
our models. The reason of this insensitivity of optical nebular
spectra to the slope of the mass function is the relatively small
fractional range in stellar mass among the ionizing stars.

For M$_{up}=30$ \msun\/ the predictions are somewhat different.  The
scatter of the data makes it difficult to say which IMF better
represent the data, but we can exclude such a small upper mass limit
for low-metallicity \hii\/ regions.  To test the suggestion that the
upper mass limit of the IMF is lower at higher metallicity we need to
examine the abundance range Z~$>$~Z$_\odot$. Even though in general
the differences using the two upper mass limits are small, and
probably within the dispersion of the data, there is one outstanding
case, where the predictions are widely divergent: the He~I~\line5876
emission line vs \x\/ diagram (Fig.~\ref{he_cluster}). The low
He~I~\line5876/\hbeta\/ values measured at high abundance
($\simeq$~2~\zsun) are in better agreement with the lower IMF mass
cutoff, while the 100~\msun\/ upper mass limit models still provide a
good fit to the data at Z~$\simeq$~\zsun. This is consistent with the
single-star results discussed in \S~\ref{teff}, where an upper limit
for \tstar\/ around 35,000~K was established. This roughly corresponds
to an O9V spectral type star, having a mass of 25~\msun\/ (Vacca,
Garmany \& Shull 1996). An upper limit for \tstar\/ of 40,000, which
seems to be excluded, corresponds to a $\sim35$~\msun\/ stellar
mass. Correcting the He~I emission lines by different, reasonable
amounts of underlying absorption (equivalent width between zero and
one) does not modify our conclusion substantially.  The overall
variation of \tstar, of the order of 15,000~K in the 0.2--2~\zsun\/
abundance range, inferred from our radiation hardness parameters,
seems too large to be accounted for with a single upper mass limit for
star formation, considering the current models of stellar evolution at
different metallicity (see Fig.~\ref{geneva}).  If our understanding
of stellar evolution is correct, our data and models therefore suggest
an IMF biased toward a significantly lower upper mass limit for
high-metallicity (between
\zsun\/ and 2~\zsun) \hii\/ regions. Whether this is an effect of a
continuous metallicity gradient, as proposed by Shields and Tinsley
(1976), or a more or less abrupt change in the star formation
properties at high abundance, cannot be judged from our work. Our
conclusion is not necessarily in contrast with previous findings.
Most authors have concentrated on rather low abundance \hii\/ regions,
nevertheless the \tstar\/ change with metallicity that we infer from
their works is consistent with ours (see for example Melnick 1992).  A
constant upper IMF is still consistent with our data for Z~$<$~\zsun.
We must consider this result with some caution, since it is of course
not only affected by the details of the nebular models (gas physical
properties, photoionization code and atomic parameters), especially in
the critical high-abundance regime, but also by the shape of the
ionizing continuum (determined by the stellar evolutionary tracks
and/or the model atmospheres).  Uncertainties in any of these will
propagate to the final result. The adoption of different stellar
tracks, for example, can produce non-negligible effects on the
computed emission-line intensities (Garc\'{\i}a-Vargas 1996,
Stasi\'{n}ska \& Leitherer 1996). The effects of the most recent
atmosphere models, including departures from LTE and stellar winds, on
the ionization structure of \hii\/ regions, need to be investigated in
detail as well.

In summary, our results imply that either: 1) The upper mass limit
decreases with increasing metal abundance; or 2) Current stellar
models underestimate the decrease in stellar temperature with
increasing abundance, at a given mass.

\section{SUMMARY}
In this paper we have presented a new spectroscopic survey of 96
\hii\/ regions contained in spiral galaxies of different Hubble
type. To interpret the measured nebular emission line ratios, two
large sets of photoionization models have been calculated: single-star
models based on the LTE Kurucz stellar atmospheres, and cluster models
in which the ionizing continuum is taken from the evolutionary
synthesis models of Leitherer \& Heckman (1995). Additional
single-star models were computed with more recent non-LTE stellar
atmosphere models.

We have mostly focused on two line diagnostics, the softness parameter
$\eta\prime$ and the He~I~\line5876/\hbeta\/ line ratio, in order to
investigate how the temperature of the ionizing stars depends on
metallicity in our sample of high-abundance \hii\/ regions.  From the
single-star models we derive a rather important dependence of the
upper bound of \tstar\/ on metallicity, from 50,000~K or more at the
low-abundance end ($\sim$~0.2~\zsun) to 35,000~K at above-solar
abundances. With our current understanding of stellar evolution at
different metallicity, this suggests a smaller upper mass limit for
star formation at high metallicity, as confirmed also by the cluster
models.

We have found that, interpreting \hii\/ regions as ionized by
single-age populations of stars, only young (1 and 2 Myr) burst models
can reproduce the observed emission line sequences. This implies
either that the importance of the W-R phase is overestimated by the
stellar evolution models, or that some physical mechanism disrupts the
\hii\/ regions in a few million years. We have shown that continuous
star formation models up to a few Myr of age can be well compared with
the data, suggesting that in real \hii\/ regions populations of stars
of different age can coexist.

The ionization parameter is found to vary approximately in the range
$-3.5 < \log U < -2.5$. Evidence for a slight trend of decreasing $U$
with increasing abundance is found. Finally, an abundance-dependent
N/O ratio is necessary to match the [NII]/[OII] model sequences to the
observations.

\acknowledgments
We thank G. Ferland for the existence of CLOUDY. We are indebted with
A. Pauldrach and R. Kudritzki for making their stellar atmosphere
models available to us. Comments from an anonymous referee helped us
to improve the manuscript.


\clearpage

\def\line{\mbox {$\lambda$}}
\def\lines{\mbox {$\lambda\lambda$}}
\def\sp{\mbox {S$^+$}}
\def\spp{\mbox {S$^{++}$}}
\def\op{\mbox {O$^+$}}
\def\opp{\mbox {O$^{++}$}}

\begin{deluxetable}{p{0.8in}l}
\tablewidth{4in}
\tablenum{1}
\tablecaption{List of abbreviations}
\tablehead{
\colhead{abbreviation}     & 
\colhead{meaning} 
}

\startdata
EW\dotfill	&	equivalent width \nl
IMF\dotfill	&	initial mass function \nl
$\alpha$\dotfill	&	{\it slope} of a power-law IMF ($dN/dm\propto m^{-\alpha}$) \nl
M$_{low}$\dotfill	&	IMF lower mass limit \nl
M$_{up}$\dotfill	&	IMF upper mass limit \nl
$Q_{H^0}$\dotfill		&	number of H ionizing photons/sec \nl
$Q_{He^0}$\dotfill		&	number of He ionizing photons/sec \nl
\tablevspace{1mm}
$\eta$\dotfill			&	$\frac{\op/\opp}{\sp/\spp}$ \nl
\tablevspace{1mm}
$\eta\prime$\dotfill		&	$\frac{\rm [OII]\lines3726,3729\;/\;\rm[OIII]\lines4959,5007} {\rm
[SII]\lines6717,6731\;/\;\rm [SIII]\lines9069,9532}$ \nl
\tablevspace{1mm}
$R_{23}$\dotfill		&	([OII]\line3727 + [OIII]\lines4959,5007)/H$\beta$ \nl
\tablevspace{1mm}
SED\dotfill	&	spectral energy distribution \nl
$U$\dotfill	&	ionization parameter \nl

\enddata
\end{deluxetable}

\begin{table}
\dummytable\label{abbreviations.table}
\end{table}

\def\etal{{\em et\thinspace al.}\ } 

\begin{deluxetable}{lcccc}
\tablenum{2}
\tablecaption{The galaxy sample}

\tablehead{
\colhead{Galaxy}     & 
\colhead{Type\tablenotemark{a}}   &
\colhead{M$^{0,i}_{B_T}$\tablenotemark{a}} &
\colhead{$v_0$\tablenotemark{a}} &
\colhead{Source of} \\
\colhead{} &
\colhead{} &
\colhead{} &
\colhead{(km s$^{-1}$)} &
\colhead{Blue Spectra\tablenotemark{b}}
}

\startdata
NGC 224 \hfill 	(M31)	&	SbI-II	    &	--21.61	& --10 &	$90''$	\nl 
NGC 598 \hfill (M33)	&	Sc(s)II-III &	--19.07	& 69   &	$90''$	\nl 
NGC 628	 \hfill (M74)  &	Sc(s)I	    &	--20.87	& 861  &	MRS	\nl
NGC 1569 &	SmIV	    &	--15.34	& 144  &	$90''$	\nl
NGC 2403 &	Sc(s)III    &	--19.47	& 299  &	$90''$, MRS	\nl
NGC 2841 &	Sb	    &	--20.65	& 714  &	MMT	\nl
NGC 3031 \hfill (M81)	&	Sb(r)I-II   &	--20.75	& 124  &	MMT	\nl
NGC 5194 \hfill (M51)	&	Sbc(s)I-II  &	--20.72	& 541  &	$90''$	\nl
NGC 3310 &	Sbc(r)	    &	--19.94	& 1073 &	$90''$	\nl
NGC 3351 \hfill (M95)  &	SBb(r)II    &	--19.78	& 641  &	MMT	\nl
NGC 3368 \hfill (M96)  &	Sab(s)II    &	--20.53	& 758  &	MMT	\nl
NGC 3521 &	Sbc(s)II    &	--20.50	& 627  &	ZKH	\nl
NGC 3623 \hfill (M65)  &	Sa(s)II	    &	--20.60	& 675  &	MMT	\nl
NGC 4258 \hfill (M106) &	Sb(s)II	    &	--21.17	& 520  &	MMT	\nl
NGC 4736 \hfill (M94)  &	RSab(s)	    &	--19.93	& 345  &	MMT	\nl
NGC 4861 &	SBmIII	    &	--17.76	& 836  &	$90''$	\nl
NGC 5236 \hfill (M83)  &	SBc(s)II    &	--20.24	& 275  &	D80	\nl
NGC 5701 &	(PR)SBa	    &	--19.62	& 1424 &	MMT	\nl
NGC 6384 &	Sb(r)I.2    &	--21.40	& 1735 &	MMT	\nl
NGC 7331 &	Sb(rs)I-II  &	--21.72	& 1114 &	MMT	\nl

\enddata
\tablenotetext{a}{From Sandage \& Tammann 1987; $v_0$ corrected to H$_0$ = 75 km s$^{-1}$ Mpc$^{-1}$} 
\tablenotetext{b}{$90''$: B\&C + Steward Observatory 90-inch telescope; 
MMT: Red Channel + MMT;
MRS: McCall \etal 1985;
ZKH: Zaritsky \etal 1994; D80: Dufour \etal 1980}

\end{deluxetable}

\begin{table}
\dummytable\label{galaxies.table}
\end{table}
\newpage


\epsscale{1.2}
\begin{figure}[ht]
\plotfiddle{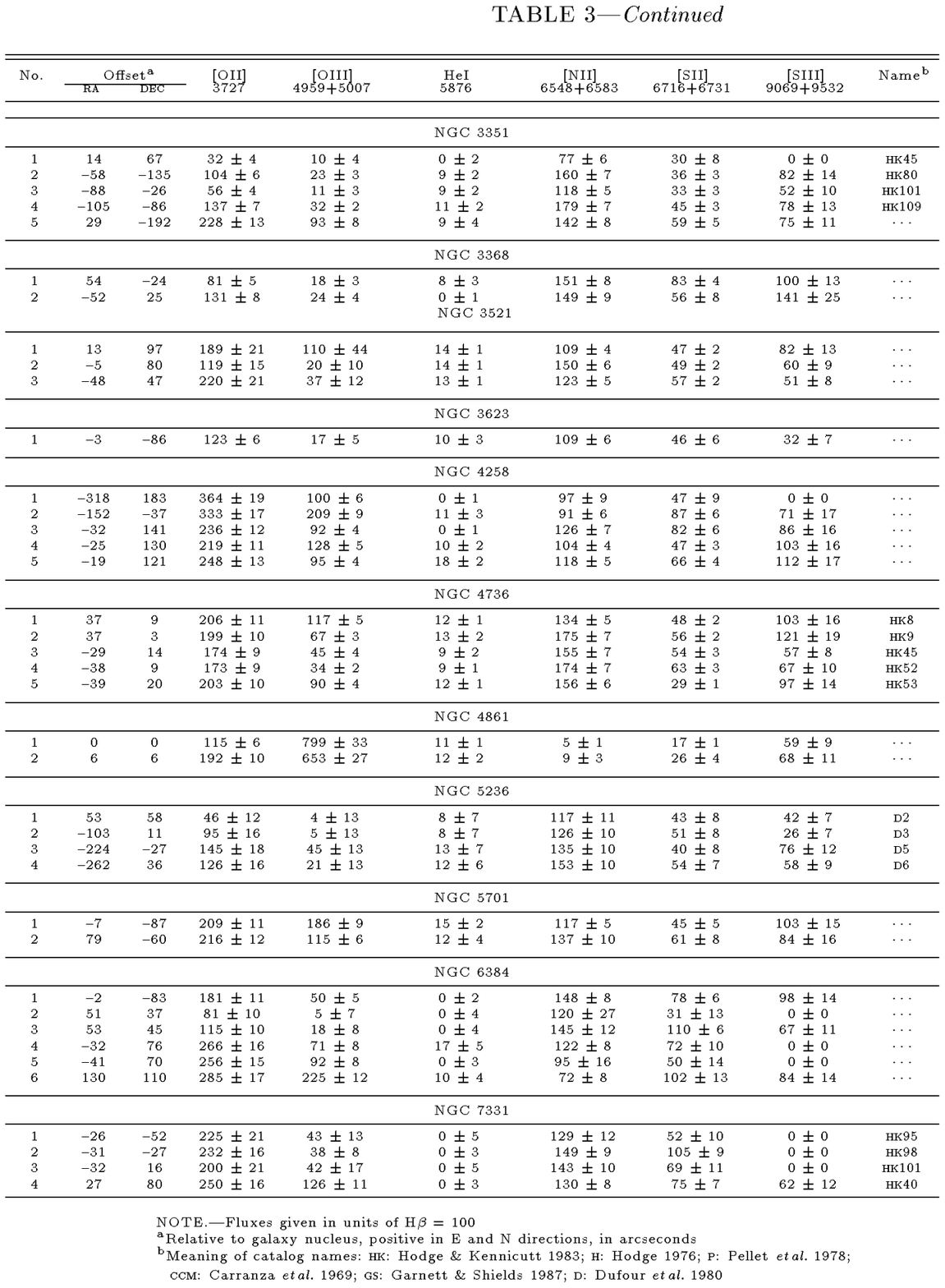}{23cm}{0}{100}{100}{-310}{-110}
\end{figure}
\epsscale{1.0}

\clearpage

\begin{deluxetable}{p{1in}c}
\tablewidth{2.8in}
\tablenum{4}
\tablecaption{Adopted solar abundances}
\tablehead{
\colhead{Element}     & 
\colhead{log abundance\tablenotemark{a}} 
}

\startdata
He\dotfill	& 	-1.00 \nl
C\dotfill	&	-3.44 \nl
N\dotfill	&	-3.95 \nl
O\dotfill	&	-3.07 \nl
Ne\dotfill	&	-3.91 \nl
Mg\dotfill	&	-5.42 \nl
Al\dotfill	&	-6.53 \nl
Si\dotfill	&	-4.75 \nl
S\dotfill	&	-4.79 \nl
Ar\dotfill	&	-5.44 \nl
Ca\dotfill	&	-6.64 \nl
Fe\dotfill	&	-5.33 \nl
Ni\dotfill	&	-6.75 \nl
Na\dotfill	&	-6.67 \nl

\enddata
\tablenotetext{a}{relative to H}
\end{deluxetable}

\begin{table}
\dummytable\label{elements.table}
\end{table}

\def\qh{\mbox {$Q_{H^0}$}}
\def\teff{\mbox {$T_{\!\!\em eff}$}}
\def\zsun{\mbox {${\rm Z_{\odot}}$}}
\begin{deluxetable}{p{1in}cccc}
\tablewidth{4in}
\tablenum{5}
\tablecaption{$Q_{He^0}$/\qh\/ from model stellar atmospheres}

\tablehead{
\colhead{\teff (K)}     & 
\colhead{2 \zsun}   &
\colhead{\zsun}   &
\colhead{0.25 \zsun}   &
\colhead{0.1 \zsun}   
}

\startdata
60,000\dotfill	& 0.35 & 0.38 & 0.45 & 0.48 \nl
50,000\dotfill  & 0.26 & 0.28 & 0.32 & 0.34 \nl
45,000\dotfill  & 0.20 & 0.22 & 0.25 & 0.26 \nl
40,000\dotfill  & 0.10 & 0.10 & 0.10 & 0.11 \nl
35,000\dotfill  & 0.01 & 0.01 & 0.01 & 0.01 \nl

\enddata
\end{deluxetable}

\begin{table}
\dummytable\label{qratio.table}
\end{table}

\def\qh{\mbox {$Q_{H^0}$}}
\def\teff{\mbox {$T_{\!\!\em eff}$}}
\def\zsun{\mbox {${\rm Z_{\odot}}$}}
\begin{deluxetable}{p{1in}cccc}
\tablewidth{4in}
\tablenum{6}
\tablecaption{$Q_{He^0}$/\qh\/ from cluster models}
\tablehead{
\colhead{age (Myr)}     &
\colhead{2 \zsun}   &
\colhead{\zsun}   &
\colhead{0.25 \zsun}   &
\colhead{0.1 \zsun}
}

\startdata
\sidehead{M$_{up}=100$ \hspace{1cm} $\alpha=2.35$} 
\tablevspace{-1mm}
\cline{1-2} 
\tablevspace{3mm}
1\dotfill  & 0.19 & 0.22 & 0.30 & 0.35 \nl
2\dotfill  & 0.11 & 0.15 & 0.24 & 0.29 \nl
3\dotfill  & 0.27 & 0.22 & 0.13 & 0.16 \nl
4\dotfill  & 0.29 & 0.21 & 0.18 & 0.10 \nl
5\dotfill  & 0.29 & 0.31 & 0.14 & 0.05 \nl
6\dotfill  & 0.24 & 0.25 & 0.01 & 0.03 \nl
\tablevspace{4mm}

\sidehead{M$_{up}=30$ \hspace{1cm} $\alpha=2.35$}
\tablevspace{-1mm}
\cline{1-2}
\tablevspace{3mm}
1\dotfill  & 0.07 & 0.08 & 0.12 & 0.15 \nl
2\dotfill  & 0.05 & 0.07 & 0.11 & 0.14 \nl
3\dotfill  & 0.03 & 0.05 & 0.09 & 0.13 \nl
4\dotfill  & 0.01 & 0.03 & 0.06 & 0.11 \nl
5\dotfill  & 0.18 & 0.01 & 0.03 & 0.06 \nl
6\dotfill  & 0.24 & 0.20 & 0.01 & 0.03 \nl
\tablevspace{4mm}

\sidehead{M$_{up}=100$ \hspace{1cm} $\alpha=3.30$}
\tablevspace{-1mm}
\cline{1-2}
\tablevspace{3mm}
1\dotfill  & 0.17 & 0.19 & 0.26 & 0.30 \nl
2\dotfill  & 0.10 & 0.13 & 0.21 & 0.26 \nl
3\dotfill  & 0.18 & 0.15 & 0.12 & 0.16 \nl
4\dotfill  & 0.23 & 0.15 & 0.13 & 0.10 \nl
5\dotfill  & 0.25 & 0.25 & 0.10 & 0.05 \nl
6\dotfill  & 0.20 & 0.21 & 0.01 & 0.03 \nl

\enddata
\end{deluxetable}

\begin{table}
\dummytable\label{qratioclusters.table}
\end{table}


\newpage

\begin{figure}[ht]
\plotone{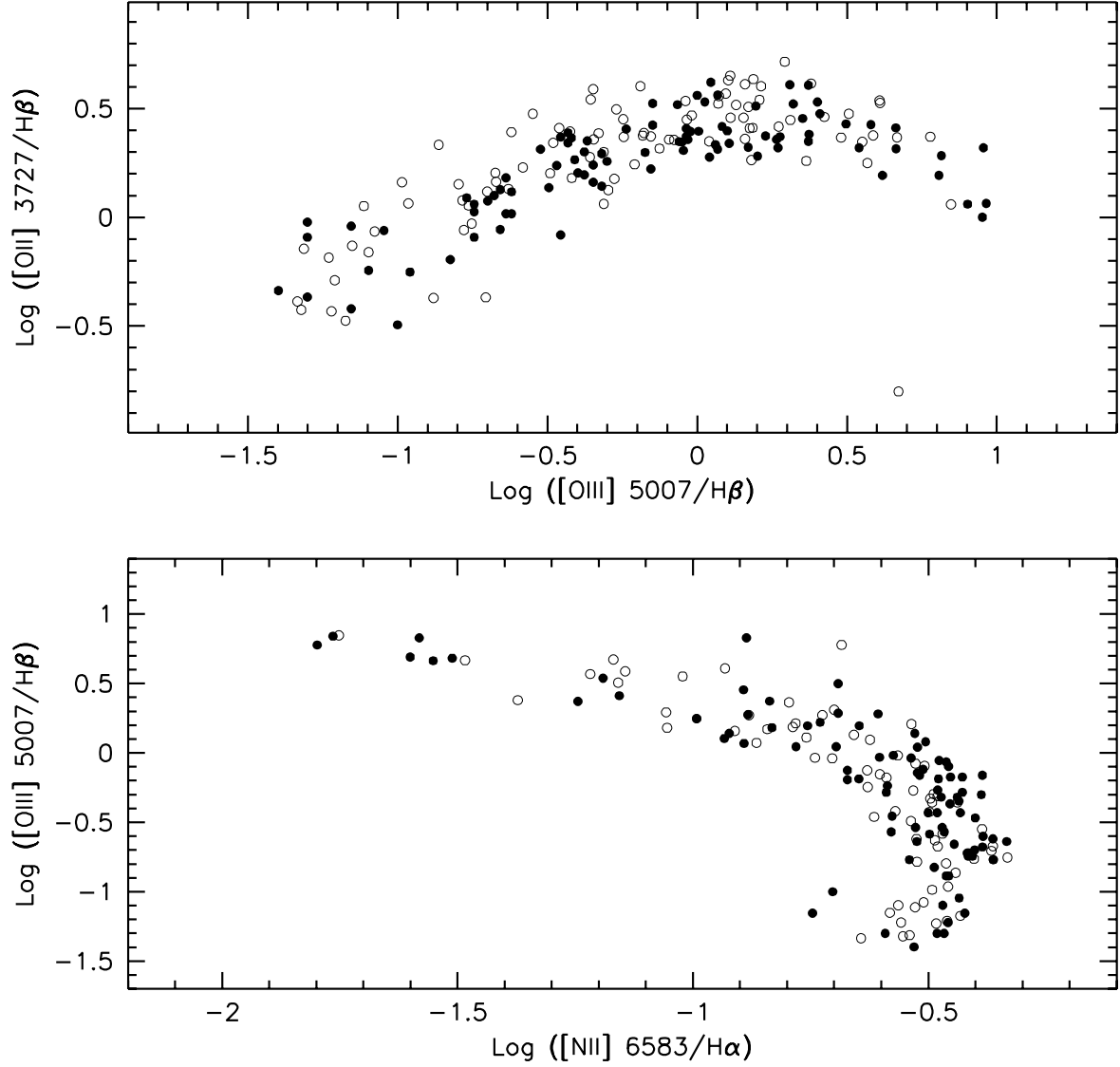}
\caption{Excitation sequences in terms of ({\em a}) [OII]/\hbeta\/ vs 
[OIII]/\hbeta\/ and ({\em b}) [OIII]/\hbeta\/ vs [NII]/\halpha. Filled
circles indicate observations from this work, open circles are data
from McCall \etal (1985).
\label{mrs}} 
\end{figure}

\begin{figure}
\plotone{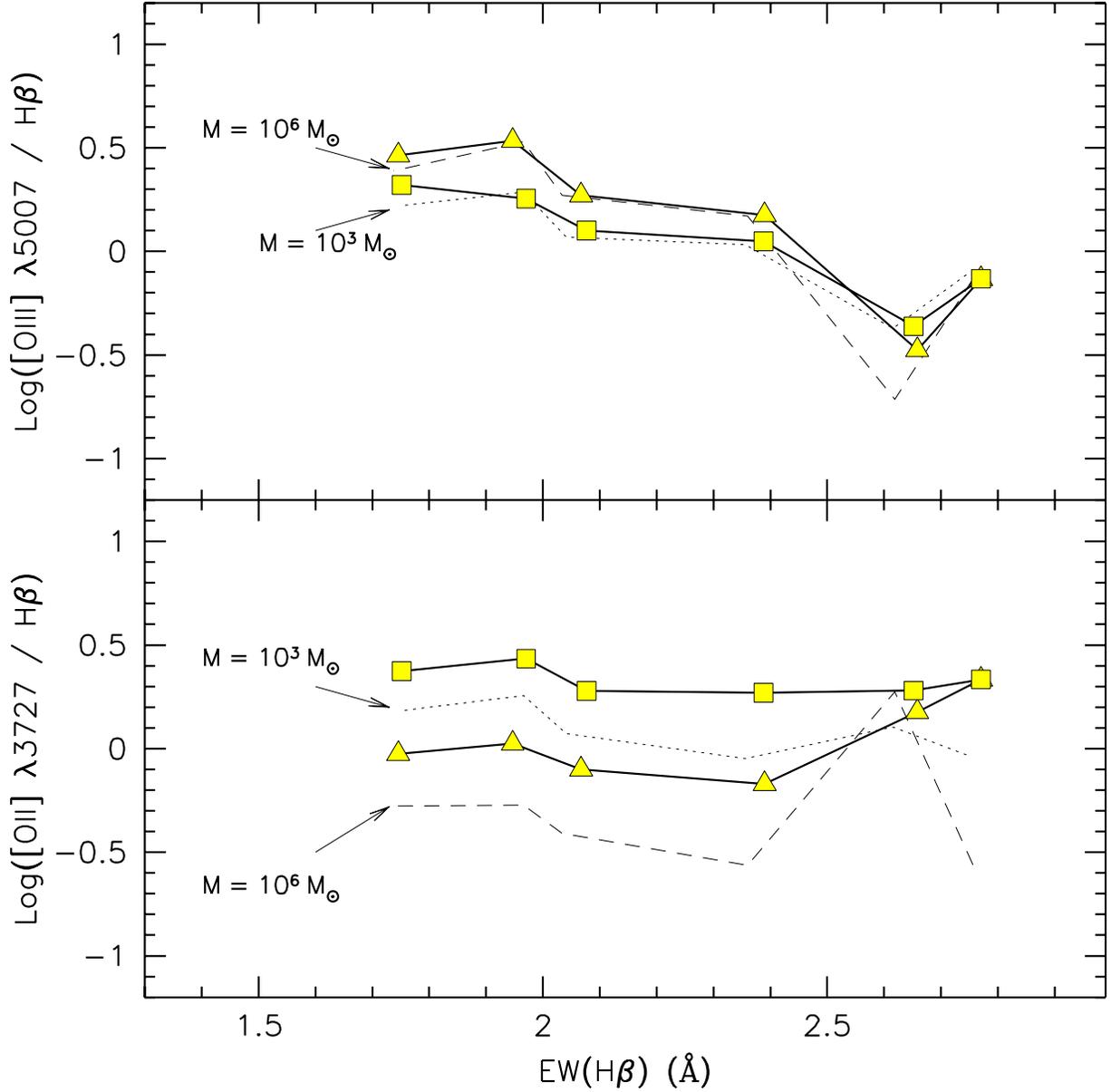}
\caption{Comparison with the models of Stasi\'{n}ska \& Leitherer
(1996). The latter are shown for an ionizing cluster mass M~=~10$^3$ and
M~=~10$^6$
\msun, with dotted and dashed lines, respectively. Our
models for $\log U\,=\,-2$ and $\log U\,=\,-3$ (calculated for
n~=~10~cm$^{-3}$) are shown by the continuous lines connecting
triangle and square symbols, respectively. All models are for solar
composition. {\em (Top)} [OIII]\line5007/\hbeta\/ vs \ew. {\em (Bottom)}
[OII]\line3727/\hbeta\/ vs \ew.
\label{comp}} 
\end{figure}

\begin{figure}
\plotone{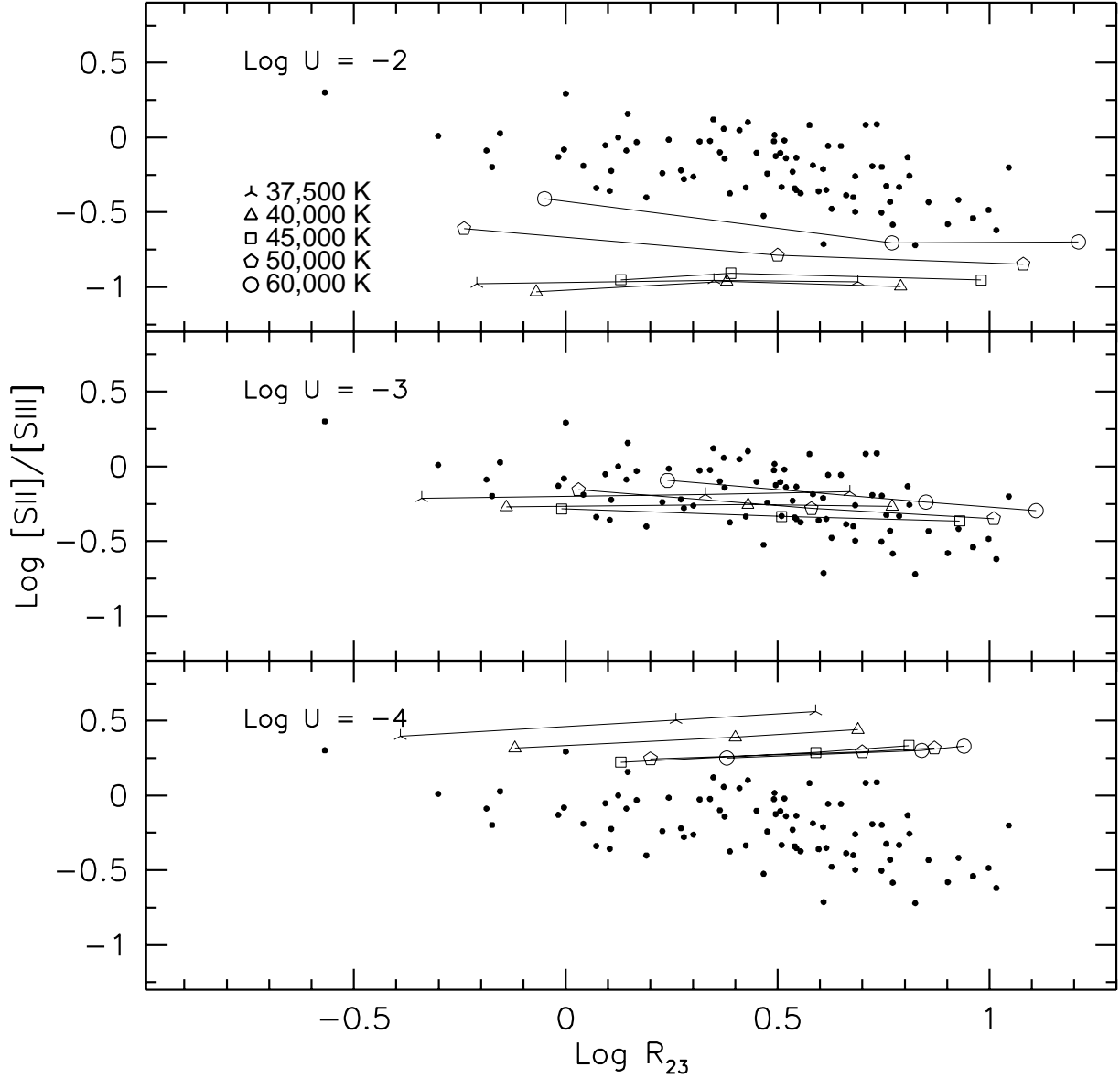}
\caption{Single-star model results for [SII]/[SIII] vs Log~\x\/, for $\log
U=-2$ (top), $\log U=-3$ (middle) and $\log U=-4$ (bottom).  The lines
connect points at three metallicity values (0.25 \zsun, \zsun\/ and 2
\zsun).  The stellar \teff\/ is coded as shown in the top panel. 
The dots are the observed values.
\label{siisiii_r23k}} 
\end{figure}

\begin{figure}
\plotone{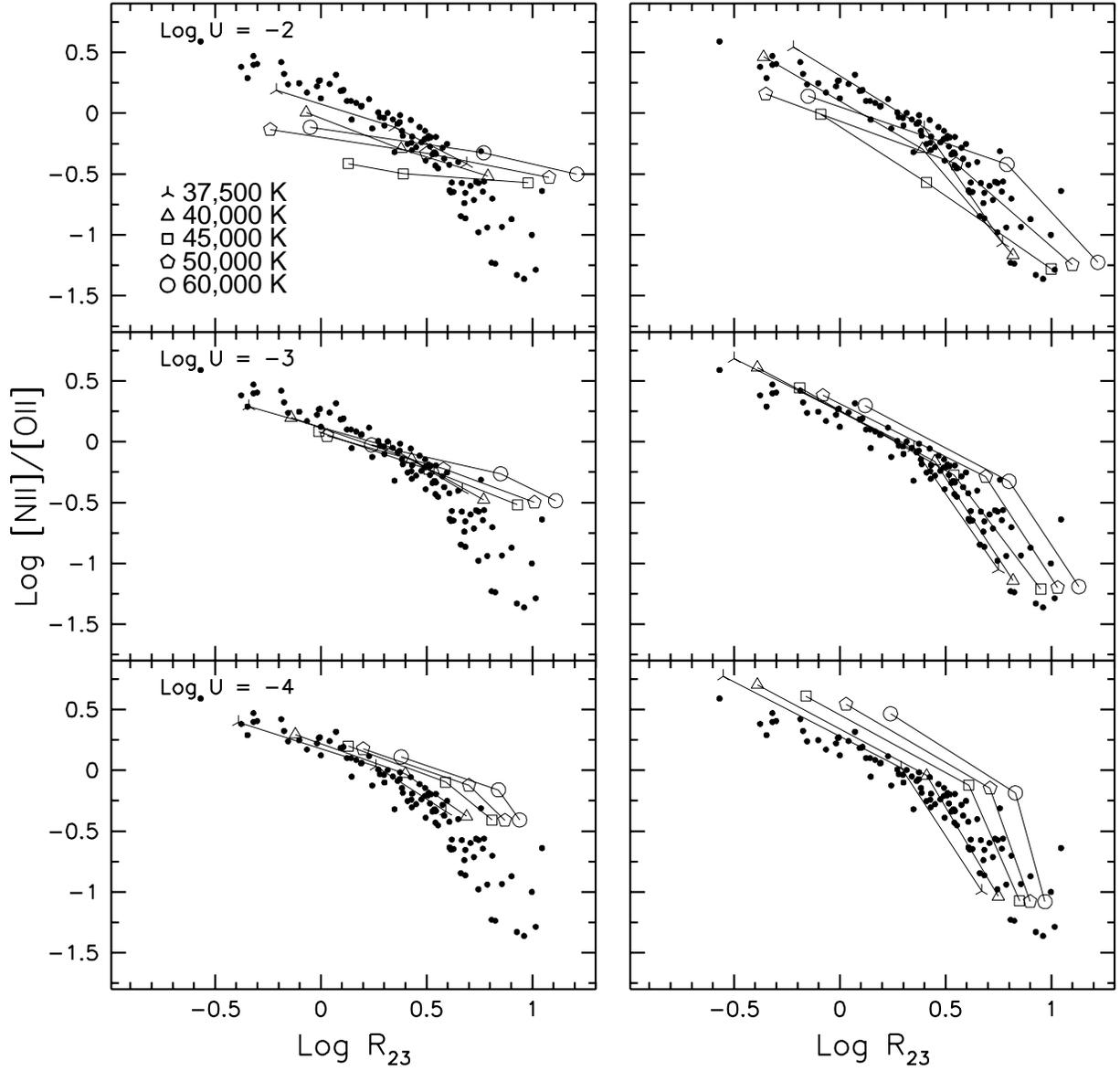}
\caption{The [NII]/[OII] vs \x\/ diagnostic diagram from models
without (left) and with (right) a metallicity-dependent N/O ratio.
Symbols as in Figure~\protect\ref{siisiii_r23k}.
\label{no}}
\end{figure}

\begin{figure}
\plotone{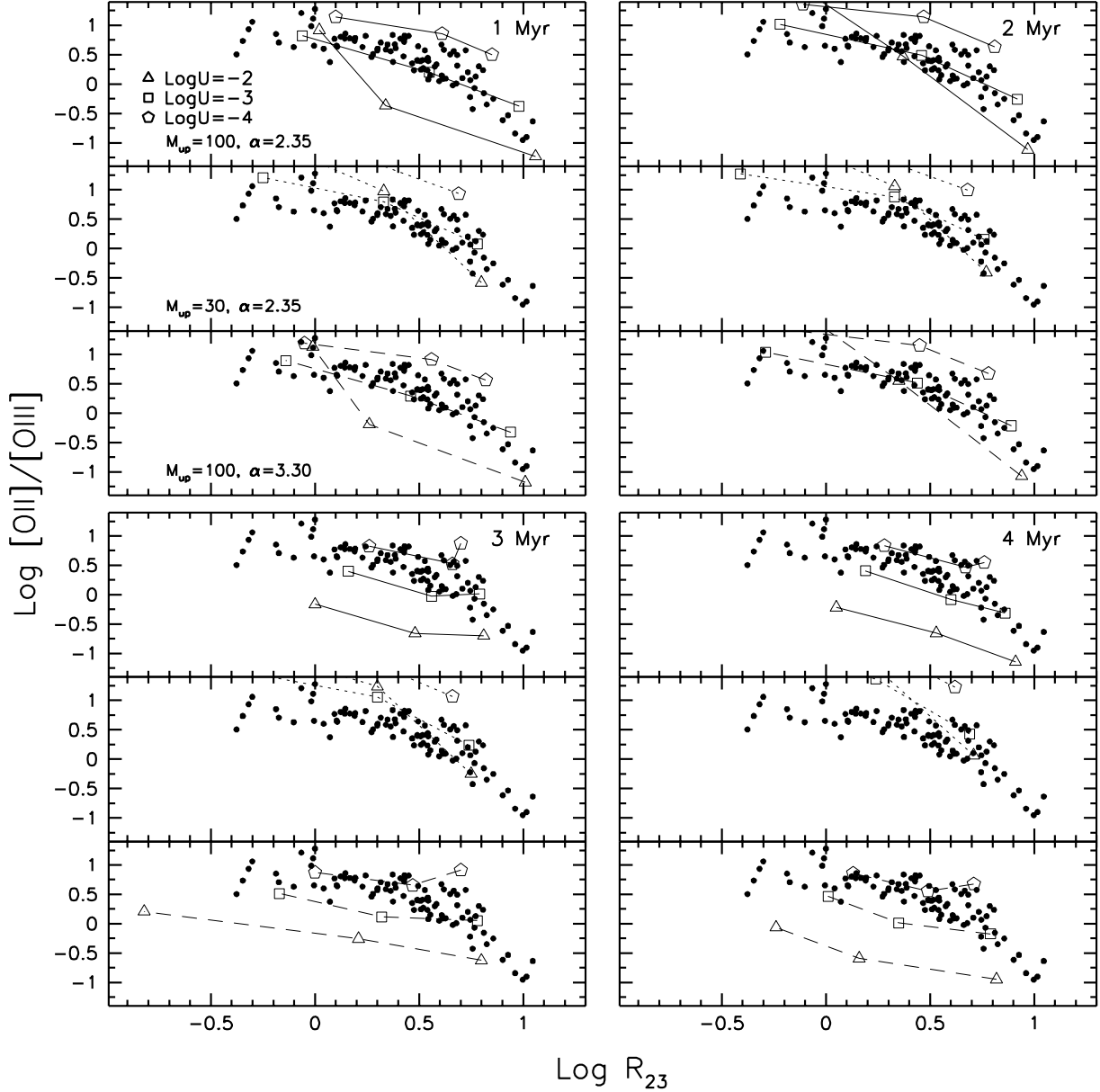}
\caption{Cluster model results for [OII]/[OIII] vs Log~\x\/. 
The two top panels refer to ages 1 and 2 Myr, the two lower panels to
ages 3 and 4 Myr. Each of these panels gives results for the Salpeter
(top), M$_{up}=30$
\msun\/ (middle), and $\alpha=3.30$ (bottom) IMFs. The lines 
connect points at three metallicity values (0.25 \zsun,
\zsun\/ and 2 \zsun). The ionization parameter is coded as 
shown in the first panel. The dots are the observed values.
\label{oiioiii_r23b}} 
\end{figure} 

\begin{figure}
\plotone{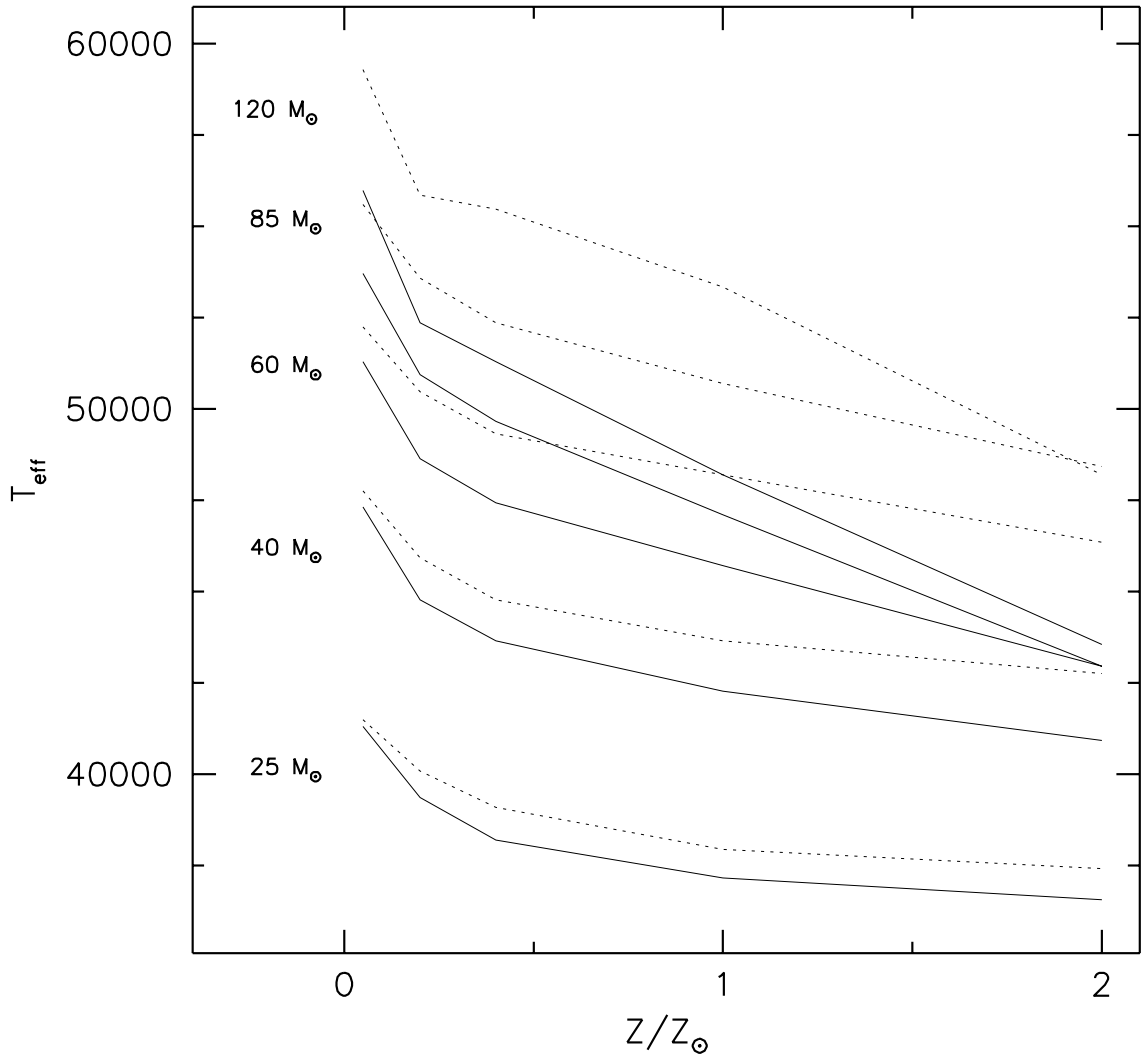}
\caption{The metallicity-\teff\/ relation predicted by the Geneva
stellar evolutionary models. The dotted lines refer to ZAMS models,
the continuous line to 1 Myr models. The stellar mass is indicated on
the left.
\label{geneva}}
\end{figure}

\clearpage

\begin{figure}
\plotone{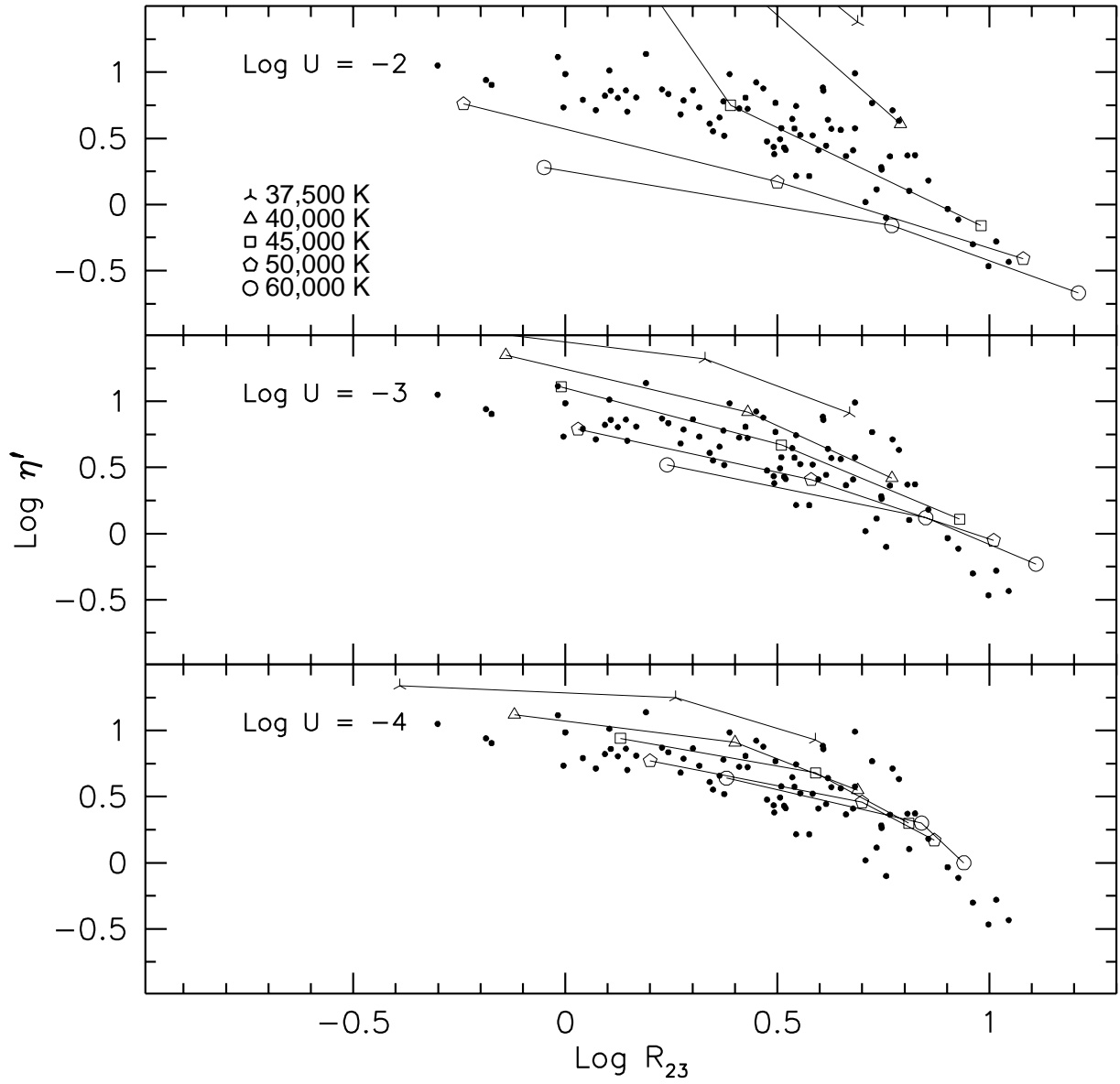}
\caption{Single-star (Kurucz atmospheres) model results for 
$\eta\prime$ vs Log~\x\/.
\label{eta_r23k}} 
\end{figure}

\clearpage

\begin{figure}
\plotone{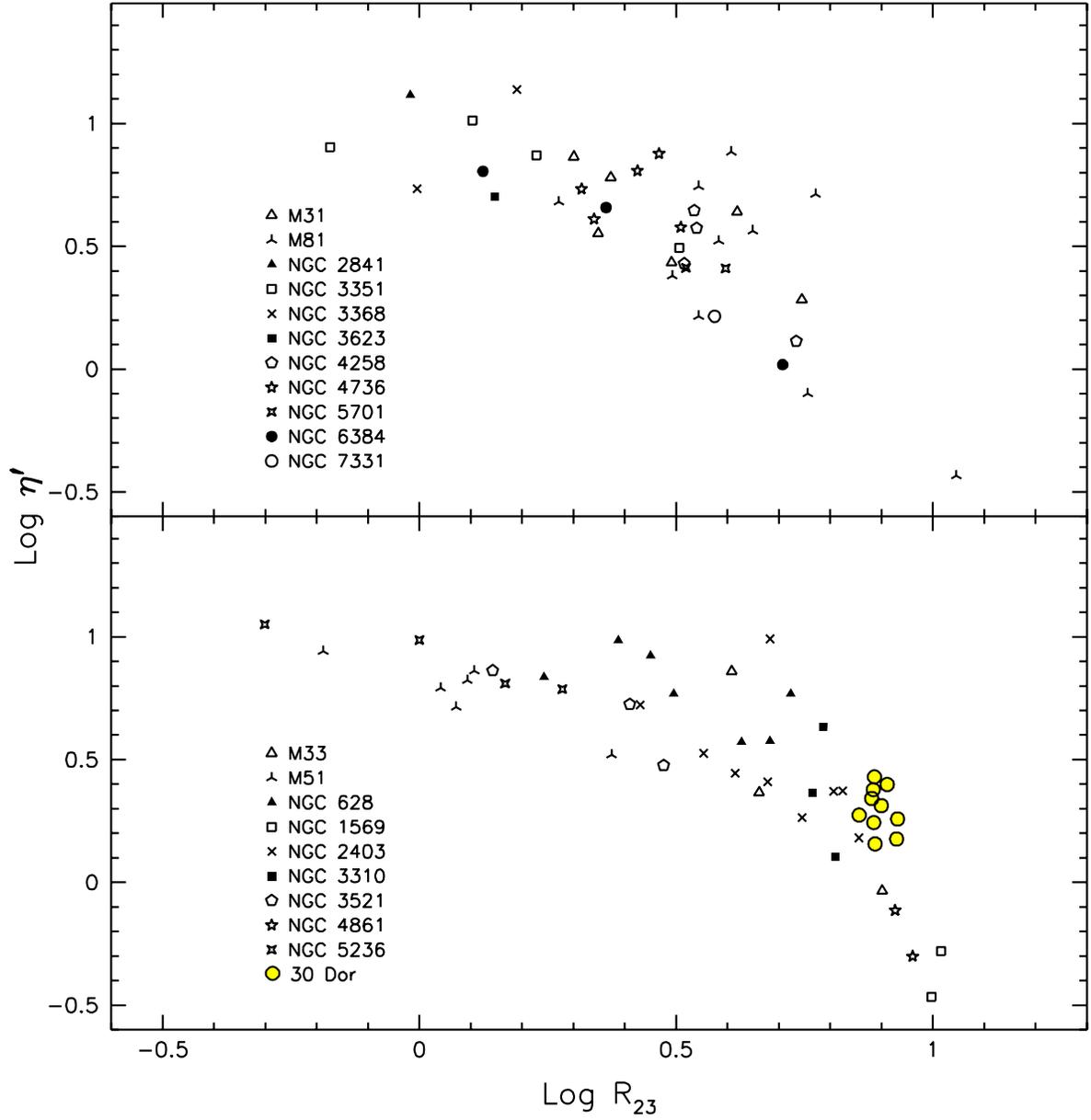}
\caption{Identification of the \hii\/ region parent galaxies 
in the $\eta\prime$~--~\x\/ diagram. Early-type (Sa--Sb) and
late-type (Sbc--Sm) spirals are shown in the upper and lower half of
the figure, respectively. Spatially resolved observations of 30 Dor
are included for comparison.
\label{labelgalaxy}} 
\end{figure}

\begin{figure}
\plotone{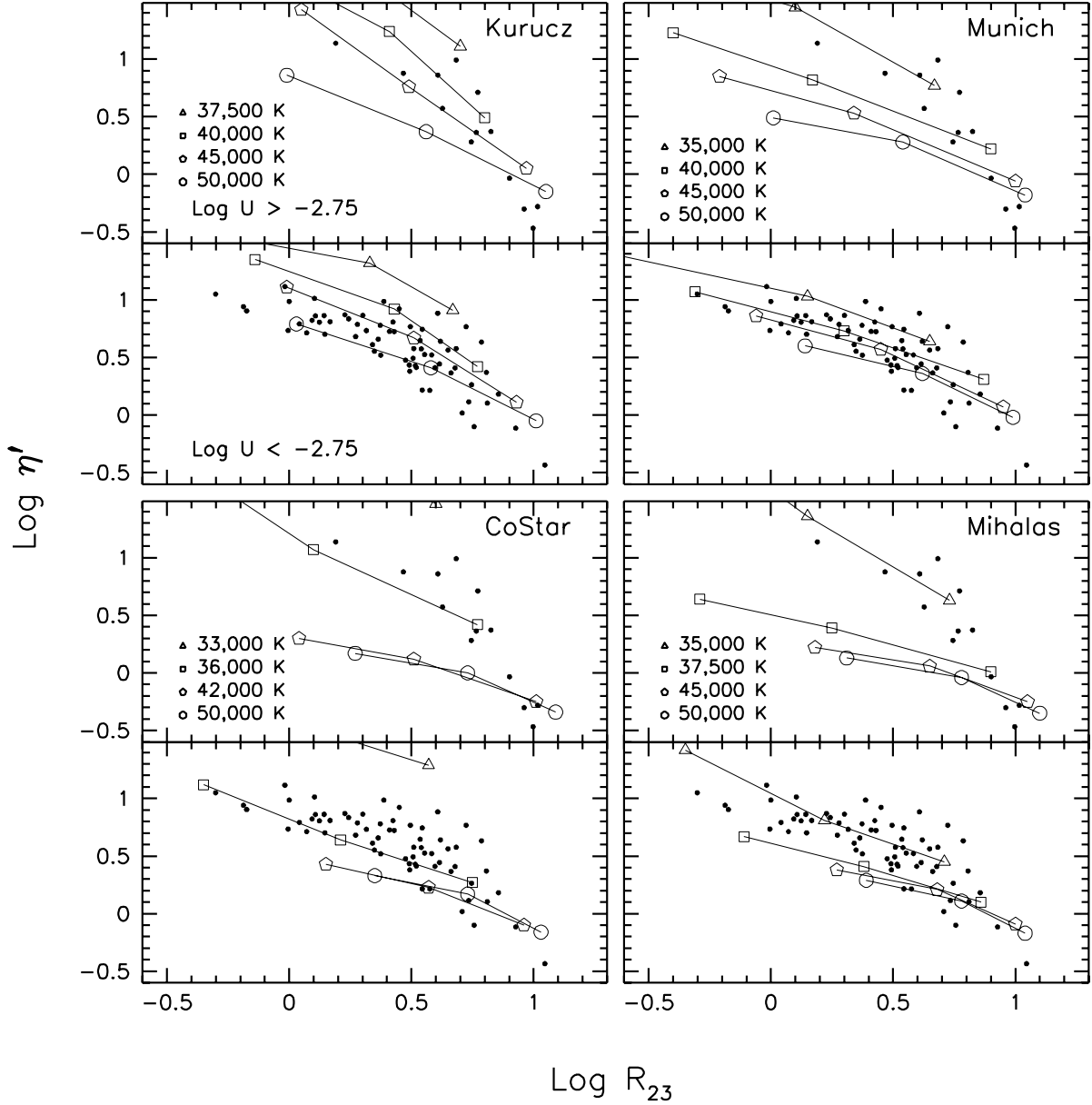}
\caption
{The nebular models based on the four theoretical stellar atmospheres
discussed in the text are compared in these $\eta\prime$ vs Log~\x\/
diagrams. For each of the four sets of photoionization models the data
points shown for comparison have been divided according to the
inferred value of $\log U$, according to the [SII]/[SIII] ratio and
the models based on the Kurucz atmospheres: $\log U > -2.75$ in the
upper half, and $\log U < -2.75$ in the lower half of each of the four
panels.  The corresponding nebular models are calculated for $\log U =
-2.5$ and $\log U = -3.0$, respectively.
\label{logn.4}} 
\end{figure}

\begin{figure}
\plotone{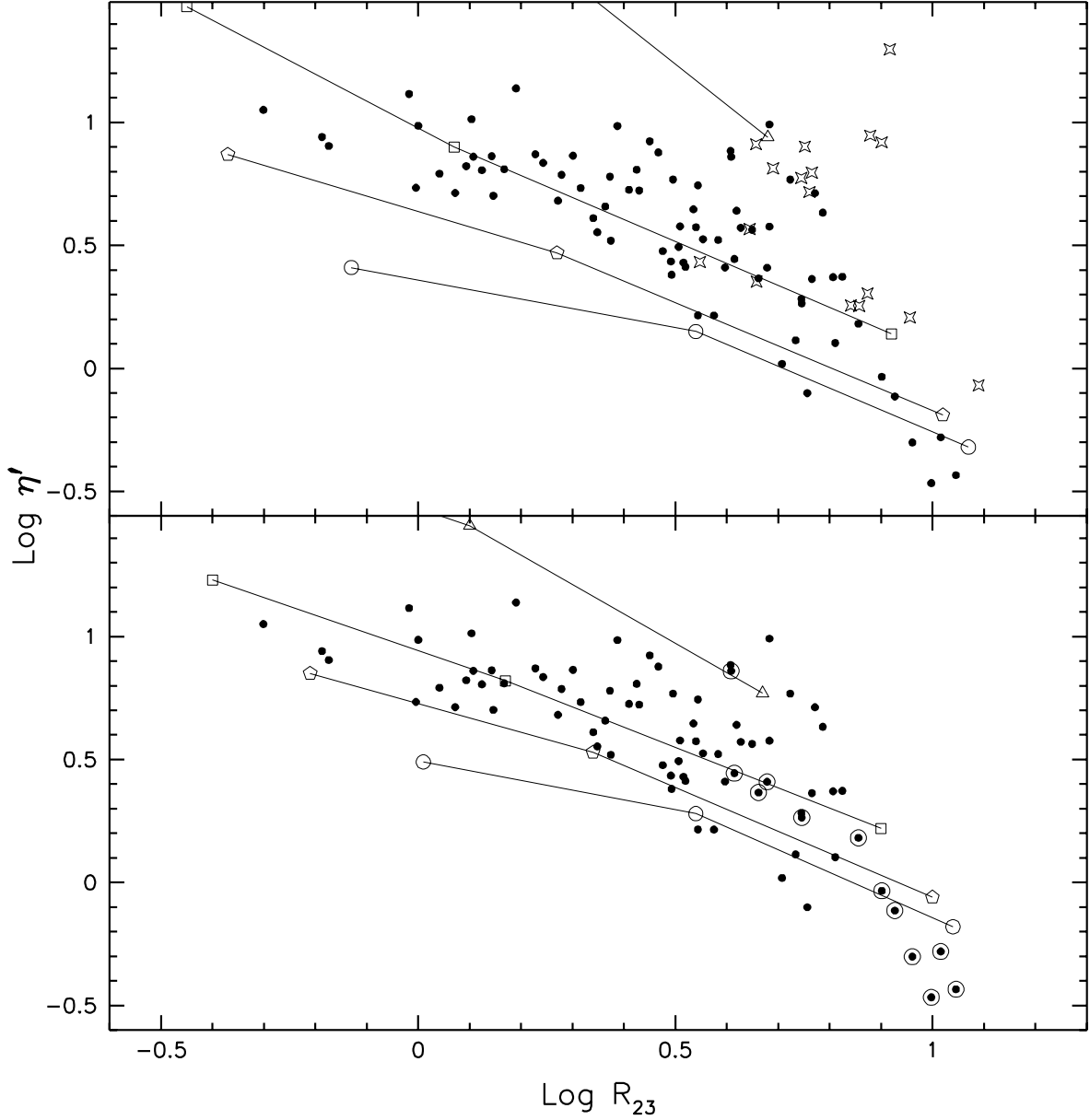}
\caption
{{\em (Top)} The open star symbols mark the position in the
$\eta\prime$--\x\/ diagram of a few galactic and Magellanic Cloud
\hii\/ regions having 0.2~$<$~Z/\zsun~$<$~0.5 (from Dennefeld \&
Stasi\'{n}ska 1983). Models calculated for $\log U = -2.0$.  {\em
(Bottom)} The objects in our sample that have O/H determined from
electron temperature measurements are marked by the larger symbols in
this diagram. Models calculated for $\log U = -2.5$.
\label{dennefeld}} 
\end{figure}

\clearpage

\begin{figure}
\plotone{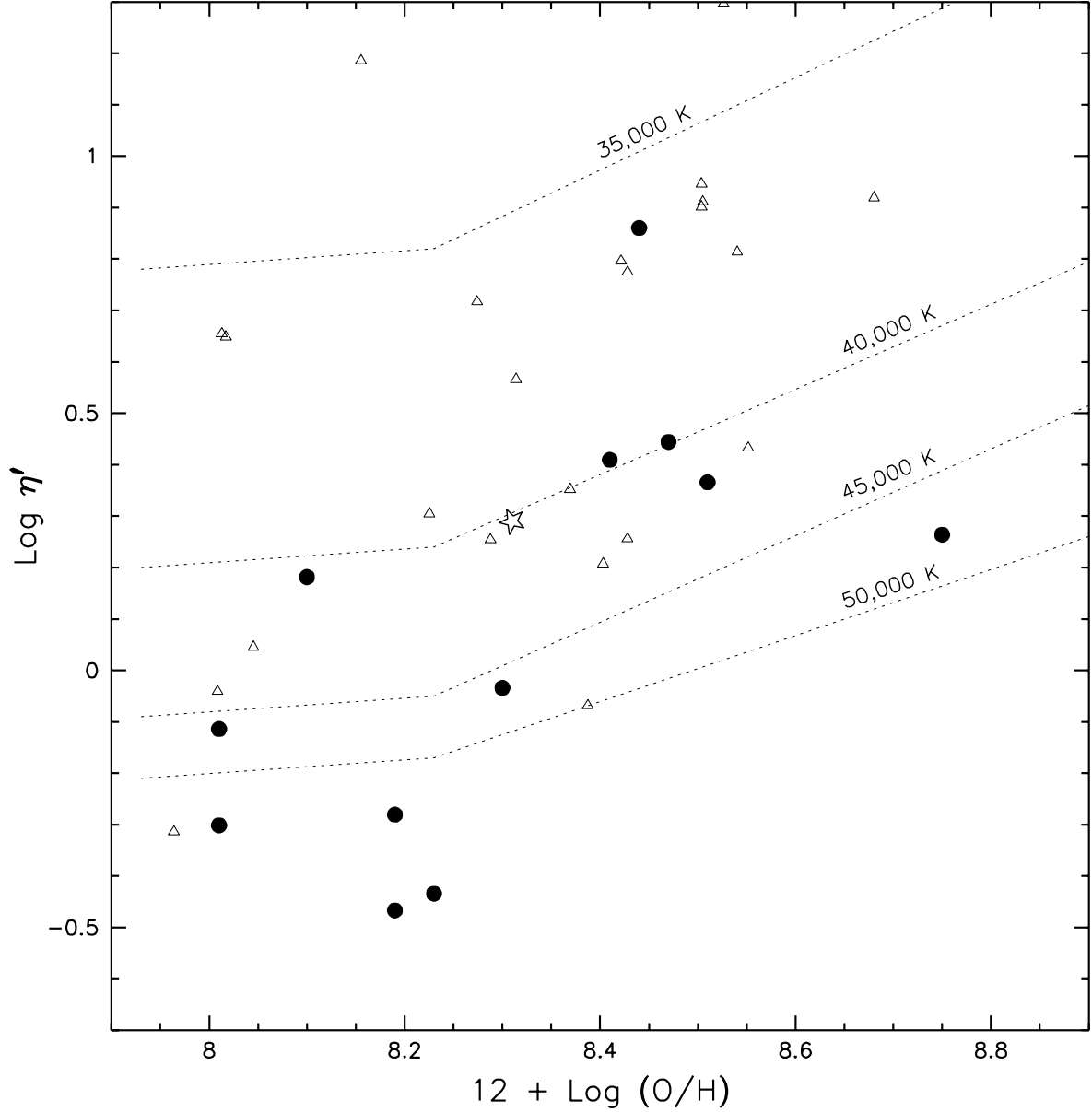}
\caption{The \tstar\/ dependence on abundance as determined using
$\eta\prime$. The data included are galactic and Magellanic Cloud
\hii\/ regions from Dennefeld \& Stasi\'{n}ska (1983; triangles),
\hii\/ regions from our extragalactic sample having O/H determinations
(solid dots), and the average of spatially resolved observations of 30
Dor (Kennicutt \etal, in preparation; open star). The models
superposed are based on the Munich group atmosphere models.
\label{oh}} 
\end{figure}

\clearpage

\begin{figure}
\plotone{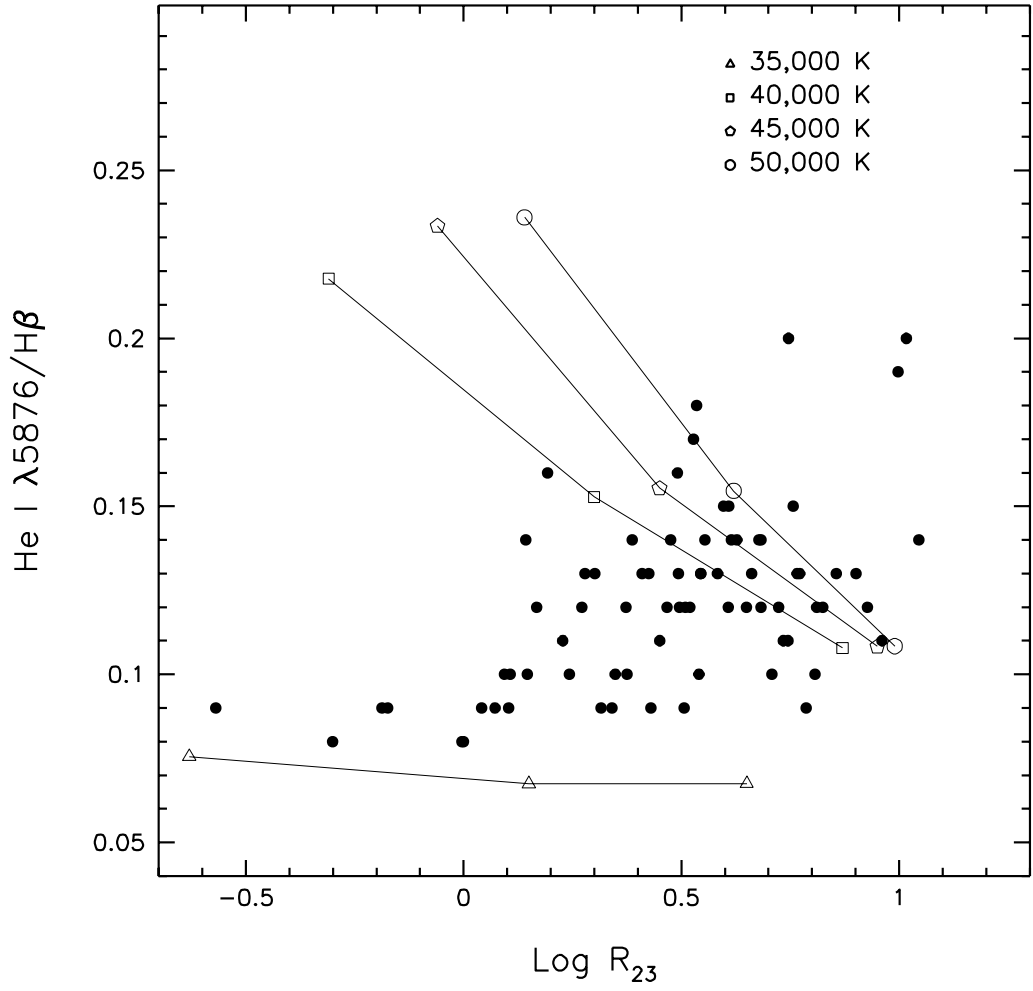}
\caption{Observations and single-star (Munich group atmospheres)
nebular models for the He~I~\line5876/\hbeta\/ diagnostic. 
\label{hei}} 
\end{figure}

\begin{figure}
\plotone{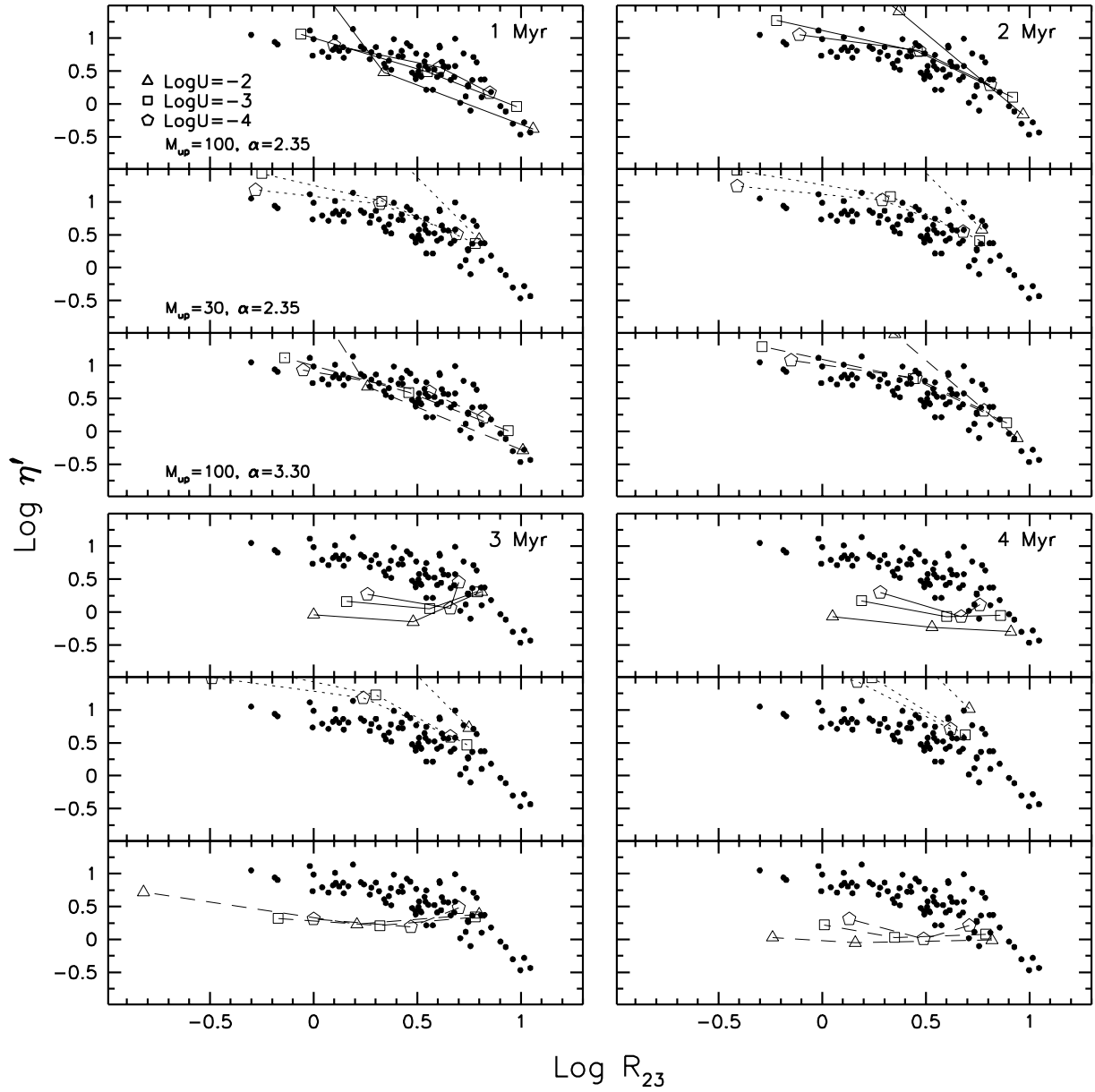}
\caption
{Cluster model results for $\eta\prime$ vs Log~\x\/.  See
Figure~\protect\ref{oiioiii_r23b} for symbol explanations.
\label{eta_r23b}} 
\end{figure}

\begin{figure}
\plotone{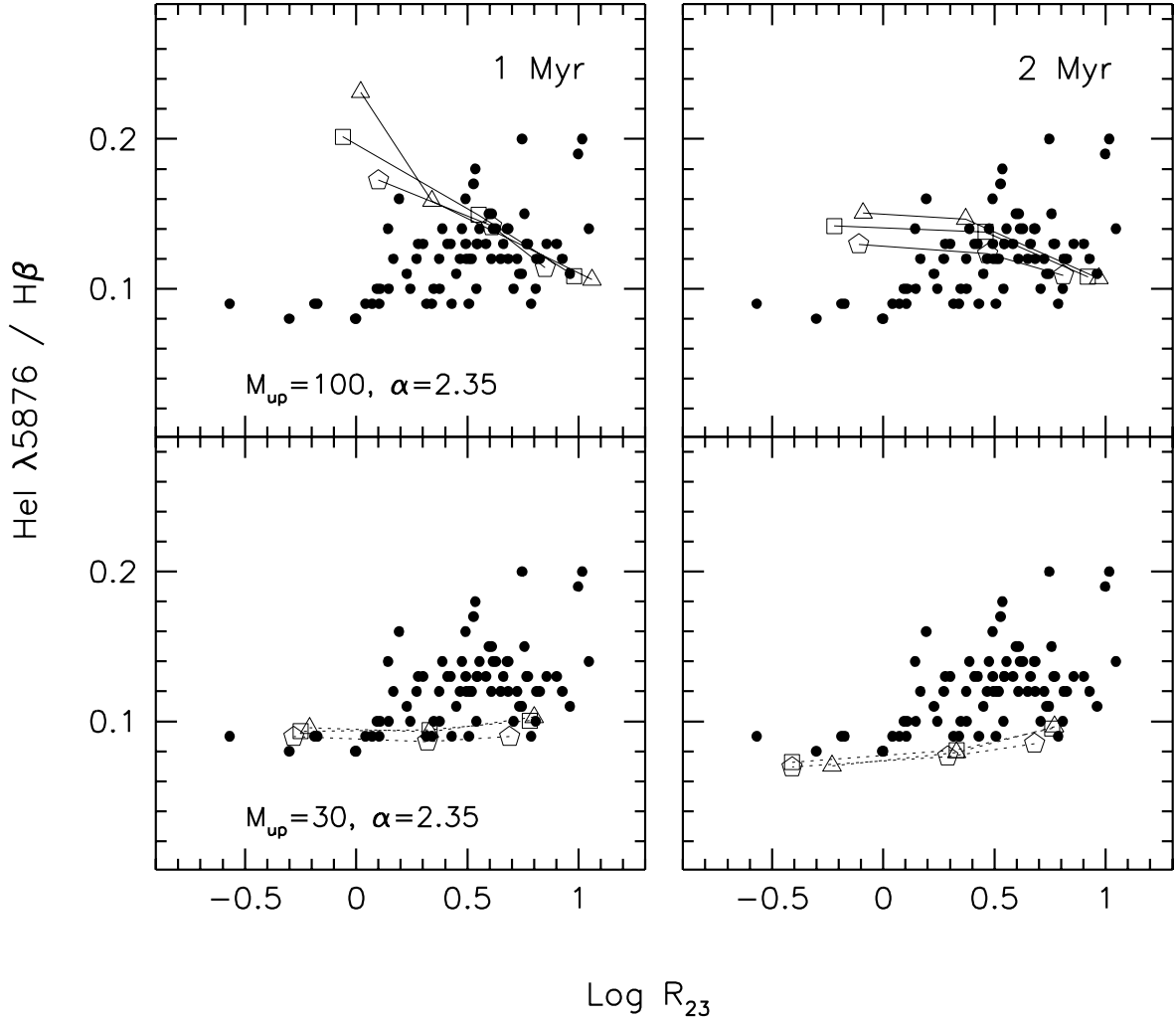}
\caption
{Observations and cluster nebular models calculated for an age of one
(left) and two (right) Myr for the He~I~\line5876/\hbeta\/
diagnostic. The top panels show models calculated with an upper mass
limit M$_{up} = 100$~\msun, the bottom panels for M$_{up} =
30$~\msun. Symbols as in Figure~\protect\ref{oiioiii_r23b}.
\label{he_cluster}} 
\end{figure}

\end{document}